\pdfoutput=1

\documentclass[11pt]{article}

\usepackage[preprint]{acl}
\usepackage{amsmath}
\usepackage{booktabs}
\usepackage{times}
\usepackage{latexsym}

\usepackage[T1]{fontenc}
\usepackage{tcolorbox}
\usepackage{enumitem}
\usepackage[utf8]{inputenc}

\usepackage{microtype}

\usepackage{multirow}
\usepackage{inconsolata}

\usepackage{graphicx}

\usepackage{color}

\usepackage{CJKutf8}
\title{PII-Bench: Evaluating Query-Aware Privacy Protection Systems}

\author{
    \textbf{Hao Shen\textsuperscript{1}},
    \textbf{Zhouhong Gu\textsuperscript{2}}
    \textbf{HaoKai Hong\textsuperscript{1}},
    \textbf{Weili Han\textsuperscript{1,3,*}}
    \\
    \textsuperscript{1}Institute of Fintech, Fudan University \\
    \textsuperscript{2}Shanghai Key Laboratory of Data Science, School of Computer Science, Fudan University \\
    \textsuperscript{3}Laboratory of Data Analytics and Security, Fudan University \\
    \small{
        \texttt{\href{mailto:email@m.fudan.edu.cn}{\{hshen22, zhgu22, hkhong23\}@m.fudan.edu.cn}} \quad
        \texttt{\href{mailto:email@fudan.edu.cn} {wlhan@fudan.edu.cn}}
    }
}

\begin{document}
\begin{CJK}{UTF8}{gbsn}
\maketitle
\begin{abstract}

The widespread adoption of Large Language Models (LLMs) has raised significant privacy concerns regarding the exposure of personally identifiable information (PII) in user prompts. To address this challenge, we propose a query-unrelated PII masking strategy and introduce PII-Bench, the first comprehensive evaluation framework for assessing privacy protection systems. PII-Bench comprises 2,842 test samples across 7 PII types with 55 fine-grained subcategories, featuring diverse scenarios from single-subject descriptions to complex multi-party interactions. Each sample is carefully crafted with a user query, context description, and standard answer indicating query-relevant PII. Our empirical evaluation reveals that while current models perform adequately in basic PII detection, they show significant limitations in determining PII query relevance. Even advanced LLMs struggle with this task, particularly in handling complex multi-subject scenarios, indicating substantial room for improvement in achieving intelligent PII masking.

\end{abstract}

\section{Introduction}

Recent years have witnessed the widespread adoption of Large Language Models (LLMs), with an increasing number of users directly interacting with these models through APIs for various tasks, ranging from daily conversations to complex analytical work (\citealp{sun2023textclassificationlargelanguage};~\citealp{yang2024zhongjing};~\citealp{WONG2023253}).
Despite the convenience these services offer, users often overlook a significant privacy risk: 
the prompts submitted to LLMs frequently contain substantial personally identifiable information (PII)~\citep{achiam2023gpt}.
Such information is vulnerable not only to interception by malicious actors during transmission~\citep{parast2022cloud} but also to potential misuse by unethical service providers who might collect and incorporate it into subsequent model training, leading to permanent privacy breaches~\citep{liu2023trustworthy}.

\begin{figure}[t]
  \includegraphics[width=\columnwidth]{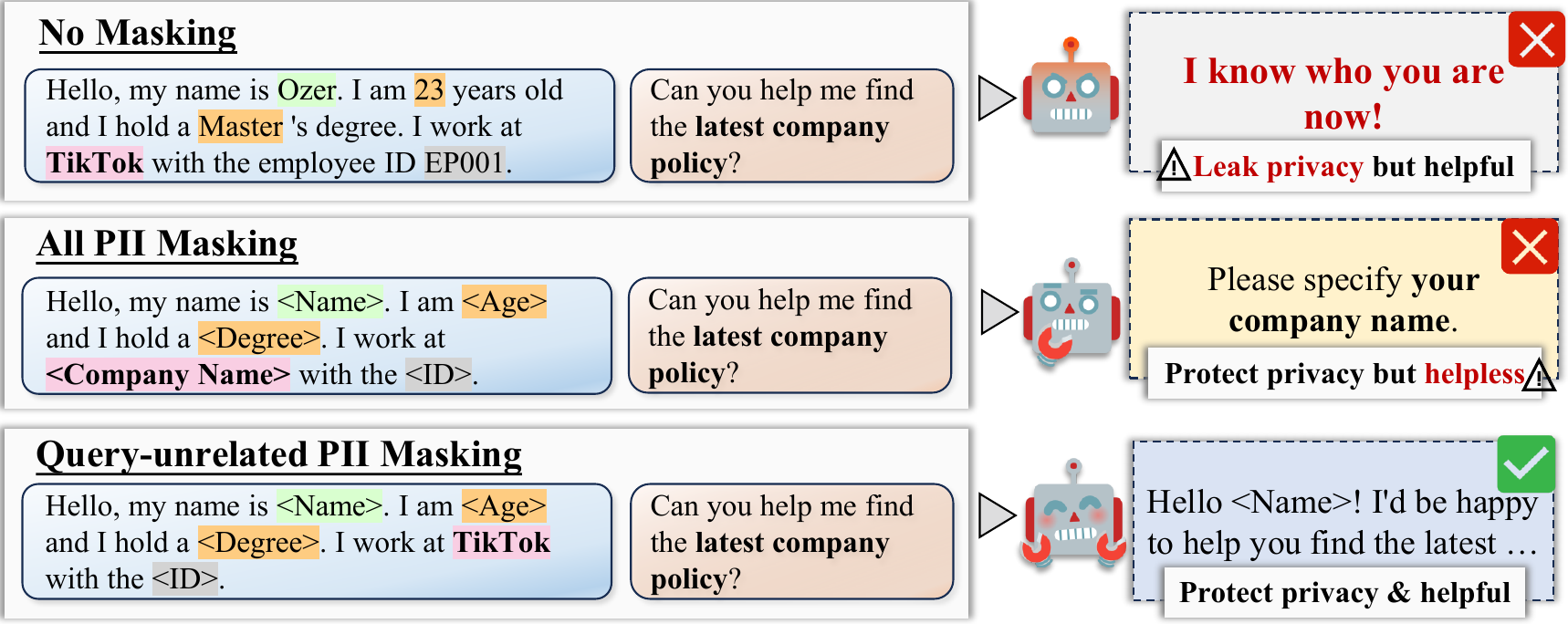}
  \caption{The overall performance of three PII Masking strategies: No Masking, All PII Masking, and Query-unrelated PII Masking.
  Effective Privacy Protection Systems are required to maintain LLMs' functionality while protect user's privacy as much as possible.
  }
  \label{fig:intro}
  \vspace{-5mm}
\end{figure}

Current practices reveal that the vast majority of users adopt a zero-protection approach when utilizing LLM services, submitting original prompts containing PII directly to the LLMs.
While an obvious protection strategy would be to mask all PII (~\citealp{nakamura2020kart};~\citealp{biesner2022anonymization};\citealp{lukas2023analyzing}), as shown in Figure~\ref{fig:intro}, this approach significantly compromises service quality.
An ideal Privacy Protection System should maintain LLMs' functionality while maximizing user privacy protection.
For instance, when a user inquires about a candidate's suitability for a senior researcher position, masking their educational background and work experience would render the LLM incapable of making an effective assessment.

This observation motivates our proposal of a query-unrelated PII masking strategy:
Masking only the PII irrelevant to user queries while retaining essential information.
In the aforementioned example, this approach would preserve the candidate's educational and professional information while masking unrelated personal details such as contact information.

The implementation of query-unrelated PII masking strategy faces two-tier challenges.
The first involves accurate identification of all PII within the prompt, serving as foundational work. 
The second requires determining the relevance of identified PII to user queries. 
While existing research has made progress in basic PII detection, systematic studies considering query relevance remain scarce.

To advance the field of privacy-preserving language models, we present PII-Bench, a comprehensive evaluation framework designed to assess Privacy Protection Systems' efficacy in preserving LLMs' core functionalities while optimizing user privacy safeguards.
PII-Bench comprising 2,842 carefully designed test samples across 7 PII types with 55 fine-grained subcategories, ranging from basic personal information to complex social relationship data. 
Each sample consists of three key components: 
(1) A user query simulating real-world information needs. 
(2) A context description containing diverse PII. 
(3) A standard answer indicating query-relevant PII and masking requirements.

Our experimental analysis reveals that while existing models, including  Bidirectional Long Short-Term Memory with Conditional Random Fields (BiLSTM-CRF)~\cite{chen2017improving}, perform adequately in basic PII detection, they demonstrate notable limitations in determining PII query relevance.
Even advanced LLMs face challenges in this task, indicating substantial room for improvement in achieving intelligent PII masking.
Despite the recent advances in model architecture and training techniques, Small Language Models (SLM) still show considerable performance gaps compared to larger LLMs, particularly in determining PII query relevance.

The primary contributions of this work include:
1. The first proposal of query-unrelated PII masking strategy, offering novel approaches to maintain LLM service quality while protecting privacy.
2. Development of PII-Bench evaluation framework, enabling systematic assessment of models' capabilities in PII identification and query relevance determination.
3. Experimental revelation of current model limitations in this task, providing direction for future research.

\section{Related Work}
\subsection{Privacy-Preserving Text Processing}
Text privacy protection has emerged as a critical challenge in natural language processing applications. \citet{papadopoulou2022neural} proposed text sanitization that combines entity detection with privacy risk assessment to guide masking decisions. \citet{shen2024fire} extended this approach with an end-to-end framework that preserves task utility during privacy protection. Exploring information preservation, \citet{meisenbacher2024just} introduced differential privacy techniques for text modification, demonstrating improved semantic retention over traditional masking methods.
While these approaches have advanced privacy protection techniques, they primarily focus on document-level sanitization without considering the dynamic nature of user interactions. Our work introduces query-aware privacy protection that adaptively balances information utility with privacy requirements.

\subsection{Query-Aware PII Detection}
Traditional PII detection has evolved from rule-based systems (\citealp{ruch2000medical};~\citealp{douglass2005identification}) to neural architectures (\citealp{deleger2013large};~\citealp{dernoncourt2017identification};~\citealp{johnson2020deidentification}), with recent work demonstrating the effectiveness of transformer-based models in identifying sensitive information \citep{asimopoulos2024benchmarking}. Large language models have shown promising results in recognizing diverse PII types (\citealp{singhal2024identifying};~\citealp{gpt4pii}), yet they treat all sensitive information with uniform importance.
Our framework introduces a novel dimension to PII detection by incorporating query relevance assessment. Rather than applying uniform protection measures, we focus on identifying which PII elements are essential for addressing user queries. This approach enables more nuanced privacy protection by distinguishing between query-related and query-unrelated sensitive information, though the actual masking or protection mechanisms are left to downstream applications.

\subsection{Privacy Protection Benchmarks}
Existing benchmarks for evaluating privacy protection methods have primarily focused on general PII detection capabilities. \citet{pilan2022text} introduced TAB, a benchmark based on legal court cases, which evaluates text anonymization performance. However, it does not assess the model's ability to distinguish query-related information. The recent work by \citet{sun2024deprompt} proposed evaluation metrics for privacy-preserving prompts, but their focus remains limited to general desensitization effectiveness. \citet{li2024llm} developed LLM-PBE to assess privacy risks in language models, though their emphasis is on model-side privacy rather than input text protection.

PII-Bench addresses these limitations by providing a comprehensive evaluation framework that assesses both PII detection accuracy and the ability to determine query-related information. This dual focus enables more realistic evaluation of privacy protection systems in interactive scenarios, where the relevance of sensitive information varies with user queries.

\section{PII-Benchmark}
\begin{figure*}[t]
\includegraphics[width=1\linewidth]{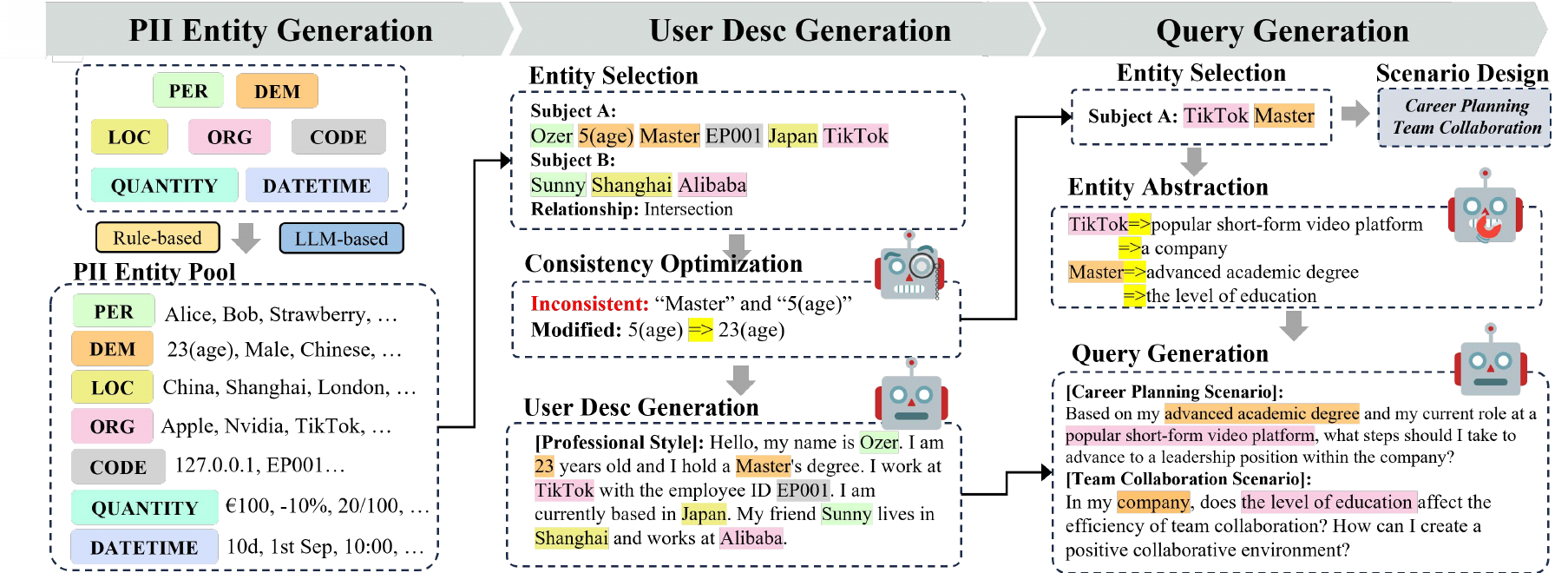}
\caption{
\textbf{PII-Bench} synthesis process consists of three main modules: (a) PII Entity Generation, (b) User Description Generation, and (c) Query Generation.
}
\label{fig:method}
\vspace{-5mm}
\end{figure*}

\begin{table}[t]
\centering
\resizebox{\columnwidth}{!}{
\begin{tabular}{cl}
\toprule
\textbf{Symbol} & \textbf{Description} \\
\hline
$p$ & A prompt consisting of a user description and a query \\
$d$ & User description containing personal information \\
$q$ & User query specifying the information need \\
$d'$ & Modified description with masked PII \\
$p'$ & Modified prompt $(d', q)$ after PII masking \\
$\mathcal{S}$ & Set of subject individuals mentioned in the description \\
$s_i$ & The $i$-th subject individual \\
$\mathcal{E}$ & Complete set of PII entities in the prompt \\
$\mathcal{E}_i$ & Set of PII entities associated with subject $s_i$ \\
$e^i_j$ & The $j$-th PII entity of subject $s_i$ \\
$\mathcal{E}_q$ & Subset of PII entities necessary for answering query $q$ \\
$\mathcal{T}$ & Set of predefined PII types \\
\bottomrule
\end{tabular}}
\caption{Notation used throughout in Task Definition.}
\label{tab:notation}
\vspace{-5mm}
\end{table}

\subsection{Task Definition}

Privacy Protection Systems target at maintaining LLM functionality while maximizing user privacy protection.
Let $p$ be a prompt consisting of a user description $d$ and a query $q$. The description $d$ contains information about multiple subject individuals $\mathcal{S} = \{s_1, ..., s_m\}$.
For each subject $s_i$, there exists an associated set of PII entities $\mathcal{E}_i = \{e^{i}_{1}, ..., e^{i}_{k}\}$. 
The complete set of PII entities in prompt $p$ is defined as $\mathcal{E} = \bigcup_{i=1}^m \mathcal{E}_i$, where each entity $e \in \mathcal{E}$ belongs to a predefined PII type from set $\mathcal{T}$ (see Appendix~\ref{sec:type}).
Let $\mathcal{E}_q \subseteq \mathcal{E}$ denote the subset of PII entities that are necessary for answering query $q$.

Based on this definition, we propose three fundamental evaluation tasks for Privacy Protection Systems:

\textbf{(1) PII Detection Task}:
Given prompt $p$, the model needs to: 
identify the minimal text spans for all PII entities $e \in \mathcal{E}$;
establish associations between each entity $e$ and its corresponding subject $s \in \mathcal{S}$;
assign the correct PII type $t \in \mathcal{T}$ to each entity $e$.

\textbf{(2) Query-Related PII Detection Task}:
Given prompt $p$, the model needs to determine the minimal subset of PII entities $\mathcal{E}_q \subseteq \mathcal{E}$. 
This subset should only contain PII entities necessary to answer query $q$, maximizing protection of non-relevant personal information.

\textbf{(3) Query-Unrelated PII Masking Task}:
This task is what we propose the optimal form of privacy protection system.
Given prompt $p$, the model should generate a modified description $d'$ where query-unrelated PII entities are masked while preserving the necessary ones. Formally, the model should identify $\mathcal{E}_q$ and generate $d'$ where all PII entities in $\mathcal{E} \setminus \mathcal{E}_q$ are masked while preserving those in $\mathcal{E}_q$. 
The masking operation should maintain text coherence and readability while ensuring effective privacy protection for non-essential personal information.
The resulting prompt $p' = (d', q)$ should enable LLMs to accurately address the query while minimizing exposure of irrelevant personal information.

\subsection{PII-Bench Construction}

Based on the task definition above, we designed an automated process for constructing the PII evaluation dataset, as illustrated in Figure~\ref{fig:method}.

\subsubsection{PII Entity Generation}
Following \citet{papadopoulou2022neural}, we expanded the PII type set $\mathcal{T}$ with 7 main types into 55 fine-grained subcategories as detailed in Table~\ref{tab:ent_gen}, employing two complementary strategies for entity generation:

(1) Rule-based Generation: Applicable for deterministic PII types with fixed formats or enumerable value sets (e.g., phone numbers, email addresses, ID numbers).
(2) LLM-based Generation: Applicable for non-deterministic PII types requiring contextual understanding and real-world knowledge (e.g., occupation descriptions, detailed addresses).

\subsubsection{User Description Generation}
\textbf{Single-Subject Description Construction}:
The construction of single-subject descriptions follows a three-stage process:

(1) Entity Selection: 
For subject $s$, randomly sample $n$ entities ($4 \leq n \leq 16$) from different PII types to construct entity set $\mathcal{E}$.
The sampling process ensures diversity of PII types while considering their natural distribution in real-world scenarios.
(2) Consistency Optimization: 
Ensure logical compatibility among entities in $\mathcal{E}$ through LLM-based verification with crafted prompts (detailed in Appendix~\ref{sec:consistency_opt}).
For example, verifies reasonable correspondence between age and educational history as shown in Figure~\ref{fig:method}.
(3) User Desc Generation: 
Selects appropriate expression styles to generate the user description.
It employs formal description formats like job resumes and employee records in professional scenarios; casual expressions like personal profiles and self-introductions in social scenarios.

\textbf{Multi-Subject Description Construction}:
The construction process for multi-subject related descriptions includes these key steps:

(1) Entity Selection: Construct relationship network $R(s_i, s_j)$ for subject pairs $(s_i, s_j)$. Relationship types include intersection relationships like colleagues and alumni, hierarchical relationships like parent-child and teacher-student, and non-intersection relationships with no direct connection.
(2) Consistency Optimization: 
Apply relationship-aware verification based on $R$ to maintain cross-subject coherence (detailed in Appendix~\ref{sec:consistency_opt}).
The optimization enforces shared attributes for intersection relationships (e.g., company address for colleagues), structural constraints for hierarchical relationships (e.g., age differences for parent-child pairs), and removes samples with irresolvable contradictions.
(3) User Desc Generation: 
This stage designs natural interaction environments matching relationship characteristics, placing subjects in realistic scenarios (like meetings, family activities) and constructing multi-party dialogue flows to reflect interactive relationships.

\subsubsection{Query Construction}
For each description $d$, we construct queries through four phases:

(1) Entity Selection: Randomly sample $k$ entities ($1 \leq k \leq 3$) from $\mathcal{E}$ to form query-relevant entity set $\mathcal{E}_q$.
(2) Scenario Design: Generate domain-specific contexts aligned with real-world applications (detailed in Appendix~\ref{sec:scenario_design}). For instance, given $\mathcal{E}_q$ containing work experience and educational background, we design scenarios such as recruitment evaluation or career planning.
(3) Entity Abstraction: Transform specific entities in $\mathcal{E}_q$ into abstract representations while preserving semantic properties.
(4) Query Generation: Synthesize natural queries $q$ by integrating abstracted entities into their corresponding scenarios.

\subsubsection{Human Verification}
All content generated by GPT-4-0806 undergoes rigorous verification by five professional annotators and the authors, focusing on:
(1) Completeness and accuracy of PII entity annotations in description $d$.
(2) Correspondence between query $q$ and query-relevant entity set $\mathcal{E}_q$.
(3) Overall semantic coherence and scenario authenticity.
Complete annotation guidelines and quality control procedures are detailed in Appendix \ref{sec:annotation}.

\subsection{Dataset Partitioning and Statistics}

\begin{figure*}[ht]
\includegraphics[width=1\linewidth]{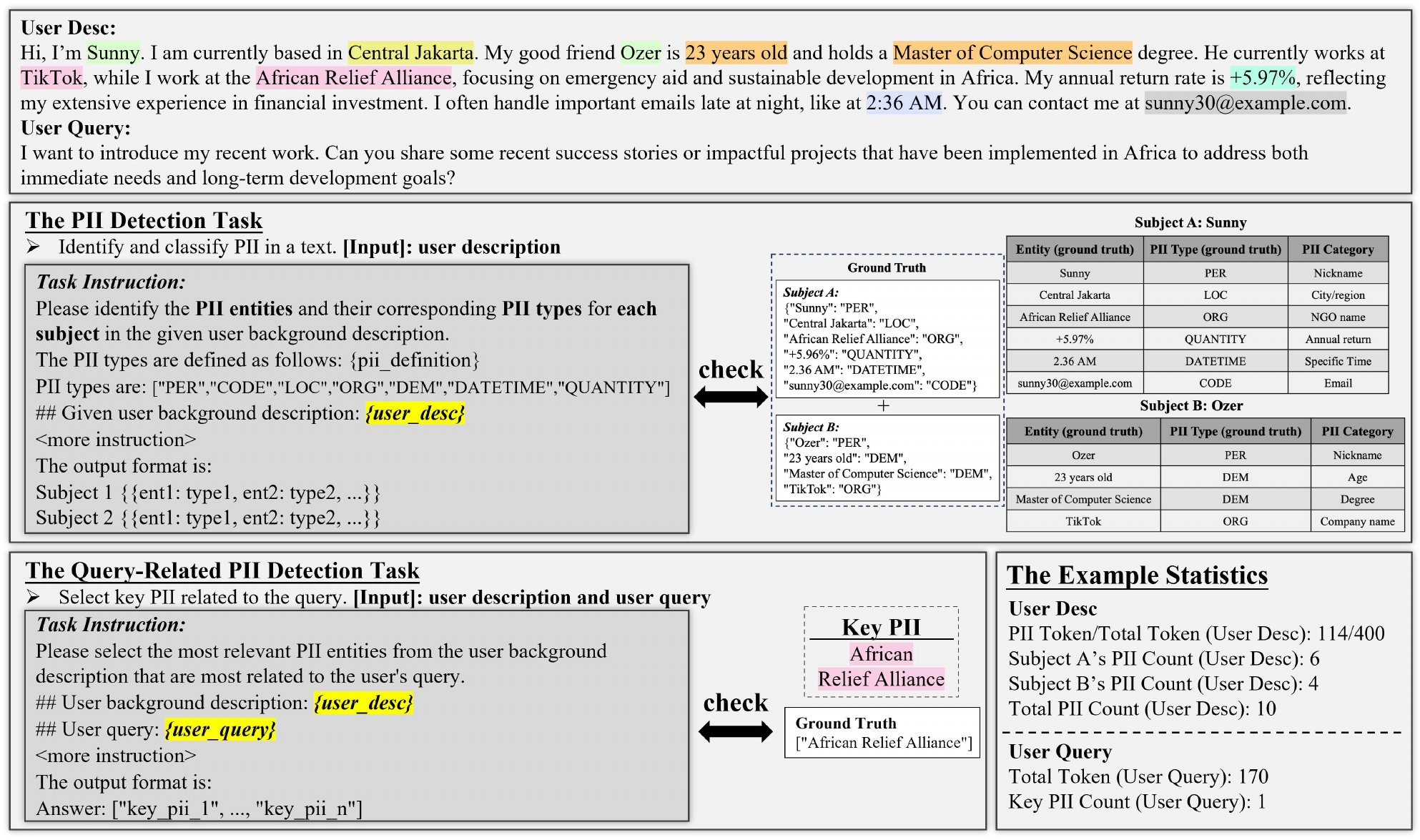}
\vspace{-5mm}
\caption{
An example from \textbf{PII-Bench}, which aims to evaluate Privacy Protection System's ability by masking maximize PII while maintaining LLM's functionality.
The evaluation is seperated by two fundamental tasks: 
(a) The PII Detection Task: Identify and classify PII entities for each subject in the prompt, with ground truth labels shown on the right side. 
(b) The Query-Related PII Detection Task: Determine which PII entities are necessary for answering the user query, enabling selective masking of irrelevant personal information. 
}
\label{fig:sample}
\vspace{-5mm}
\end{figure*}

Table~\ref{tab:stats} presents the partition and key statistics of PII-Bench, which comprises two main datasets (PII-single and PII-multi) and two specialized test sets (PII-hard and PII-distract).
Each sample follows a consistent JSON structure containing four key components:
user description, query, comprehensive PII entity annotations, and query-relevant PII labels, as illustrated in Figure~\ref{fig:sample}.

\textbf{PII-Single and PII-Multi}:
Based on the number of subjects in descriptions, the dataset is divided into two main subsets.
PII-Single contains 1,214 description-query pairs involving single subjects, focusing on model performance in handling individual information.
PII-Multi contains 1,228 description-query pairs involving multiple related subjects, evaluating model capability in handling privacy information within complex interpersonal networks.

\textbf{Test-Hard Construction}:
Select 200 challenging instances from PII-Single and PII-Multi to construct Test-Hard dataset, based on criteria including:
(1) Maximum character length of description text $d$.
(2) Highest PII entity density ($|\mathcal{E}|/|d|$).
(3) Samples with the most query-relevant entities ($|\mathcal{E}_q|$).

\textbf{Test-Distract Construction}:
Construct 200 samples simulating complex multi-user interaction scenarios.
Each sample integrates five different descriptions from PII-Single and PII-Multi, and constructs queries involving three of these descriptions based on professional networks, knowledge platforms, and community forum interaction templates.
The generation process employs dialogue flow transformation strategies to ensure natural transitions and semantic coherence, simulating real-world information interference and complex interaction patterns.

\subsection{Human Performance}


To establish a human baseline, we recruited 25 graduate students with at least two years of research experience in privacy protection and data security from top Chinese universities. 
All participants completed comprehensive training and passed a qualification test before formal evaluation (details in Appendix~\ref{sec:human_eval}).
We evaluated 400 randomly sampled instances (200 each from PII-single and PII-multi) and 100 instances from PII-distract, with each instance independently assessed by five participants.
Participants performed two sequential tasks: 
PII detection, which involved determining minimal text spans, associated subjects, and PII types for all entities in the user description, followed by query-relevant PII detection to identify entities essential for addressing the given query. 
The result of the human baseline is shown in Table~\ref{tab:human_perf}.

\begin{table}[t]
\small
\centering
\begin{tabular}{cccc}
\toprule
\textbf{Dataset} & \textbf{PII-F1} & \textbf{Query-F1} \\ \hline
PII-single & 97.2 ± 1.1 & 95.1 ± 1.3 \\ \hline
PII-multi & 95.4 ± 1.2 & 94.3 ± 1.5 \\ \hline
PII-hard & 91.3 ± 1.1 & 90.3 ± 1.2 \\ \hline
PII-distract & 92.8 ± 1.8 & 91.5 ± 2.1 \\ 
\bottomrule
\end{tabular}
\caption{Human performance in PII-Bench. PII-F1 measures accuracy in the PII detection task while Query-F1 evaluates the query-relevant PII detection task.}
\label{tab:human_perf}
\vspace{-4mm}
\end{table}

\begin{table}[t]
\centering
\resizebox{\columnwidth}{!}{
\begin{tabular}{ccccccc}
\toprule
\textbf{Name} &
  \textbf{\#Sample} &
  \textbf{Avg \#Subject} &
  \textbf{\begin{tabular}[c]{@{}c@{}}Avg \#Char\\ (Desc)\end{tabular}} &
  \textbf{\begin{tabular}[c]{@{}c@{}}Avg \#PII\\ (Desc)\end{tabular}} &
  \textbf{\begin{tabular}[c]{@{}c@{}}Avg \#Char\\ (Query)\end{tabular}} &
  \textbf{\begin{tabular}[c]{@{}c@{}}Avg \#PII\\ (Query)\end{tabular}} \\ \hline
PII-single   & 1,214          & 1.0           & 893.48           & 7.67           & 211.21          & 1.95          \\ \hline
PII-multi    & 1,228          & 2.0           & 652.65           & 13.14          & 236.21          & 2.06          \\ \hline \hline
PII-hard     & 200           & 1.5          & 778.03           & 10.60           & 222.09          & 2.10          \\ \hline
PII-distract & 200           & 7.5           & 4,403.64          & 51.08          & 859.69          & 5.82          \\ \hline \hline
\textbf{All} & \textbf{2,842} & \textbf{1.92} & \textbf{1,028.32} & \textbf{13.30} & \textbf{268.41} & \textbf{2.28} \\ 
\bottomrule
\end{tabular}
}
\vspace{-3mm}
\caption{Statistic information of PII-Bench.}
\label{tab:stats}
\vspace{-5mm}
\end{table}

\section{Experiments}
\subsection{Overall Setup}
\textbf{Traditional Model Baselines:}
We implemented \textbf{BiLSTM-CRF} as a traditional sequence labeling baseline, following the architecture proposed by \citet{huang2015bidirectional}. We trained the model using Adam optimizer with a learning rate of 1e-3 and batch size of 32 for 50 epochs on the PII-Bench training set.

\textbf{LLM Baselines:}
The evaluation encompassed both API-based and open-source large language models.
API-based models included GPT-4o-2024-0806 (\textbf{GPT4o}) \citep{gpt-4o}, Claude-3.5-Sonnet (\textbf{Claude3.5}) \citep{claude}, and DeepSeek-Chat {\textbf{DeepseekV3}} \citep{deepseek}, accessed through their respective official APIs between January 1 and February 10, 2025.
Open-source alternatives comprised Llama-3.1-70B-Instruct (\textbf{Llama3.1}) \citep{dubey2024llama}, and Qwen-2.5-72B-Instruct (\textbf{Qwen2.5}) \citep{qwen25}.

\textbf{SLM Baselines:}
To investigate scaling effects, we included two small-scale language models: Llama-3.1-8B-Instruct (\textbf{Llama3.1-SLM}) and Qwen-2.5-7B-Instruct (\textbf{Qwen2.5-SLM}).
All experiments utilized default parameters with temperature set to 0 to ensure reproducibility. Additionally, we conducted a comprehensive evaluation on smaller, deployment-ready models (0.5B-3B parameters) to assess their viability for on-device PII protection, with detailed results presented in Appendix~\ref{sec:small-models}. 

\textbf{Prompt Baselines:}
The assessment incorporated multiple prompting strategies for query-related PII detection. \textbf{Naive} inputs the user description and query. 
\textbf{Naive /w Choice} includes a list of candidate PII entities to constrain the selection space.
\textbf{Self-CoT}~\citep{wei2022chain} incorporating step-by-step reasoning prompts. 
\textbf{Auto-CoT}~\cite{autocot}, which automates the generation of chain-of-thought demonstrations through three-shot setting. 
\textbf{Self-Consistency (SC)}~\cite{sc}, which  synthesizes multiple reasoning paths to derive the final output. 
\textbf{Plan-and-Solve CoT (PS-CoT})~\cite{wang2023plan} develops a strategic plan before executing the solution process. Appendix~\ref{sec:prompt} provides details of each prompts.

\textbf{Metrics:}
The PII detection task evaluates model performance through two sets of metrics: 
\textbf{Strict-F1} measures the accuracy of subject identification, entity span detection, and PII type classification simultaneously.
\textbf{Ent-F1} focuses on entity span detection independent of subject attribution and type classification. 
For query-related detection, model performance is measured through \textbf{Precision}, \textbf{Recall}, and \textbf{F1}. 
Considering the inherent variation in entity expressions and potential partial matches, \textbf{RougeL-F} is employed for both tasks to complement the exact matching metrics. 
Detailed computation procedures are provided in Appendix~\ref{sec:metrics}. 
All reported model performance metrics are mean scores over test sets (temperature=0, top\_k=1).

\begin{table*}[htp]
\centering
\resizebox{2\columnwidth}{!}{
\begin{tabular}{ccccccccccc}
\toprule
\multicolumn{1}{c}{} & \multicolumn{2}{c}{\textbf{GPT4o}} & \multicolumn{2}{c}{\textbf{Llama3.1}} & \multicolumn{2}{c}{\textbf{Qwen2.5}} & \multicolumn{2}{c}{\textbf{Llama3.1-SLM}} & \multicolumn{2}{c}{\textbf{Qwen2.5-SLM}} \\
\multicolumn{1}{c}{\multirow{-2}{*}{\textbf{Method}}} & F1 & \multicolumn{1}{c}{RougeL-F} & F1 & \multicolumn{1}{c}{RougeL-F} & F1 & \multicolumn{1}{c}{RougeL-F} & F1 & \multicolumn{1}{c}{RougeL-F} & F1 & RougeL-F \\ \hline
\multicolumn{11}{l}{\cellcolor[HTML]{EFEFEF}\textit{Basic Method}} \\ \hline
\multicolumn{1}{l}{Naive} & 0.72 & \multicolumn{1}{c}{0.72} & 0.72 & \multicolumn{1}{c}{0.73} & 0.70 & \multicolumn{1}{c}{0.70} & 0.42 & \multicolumn{1}{c}{0.43} & 0.54 & 0.58 \\ \hline
\multicolumn{11}{l}{\cellcolor[HTML]{EFEFEF}\textit{Advanced Method}} \\ \hline
\multicolumn{1}{l}{Self-CoT} & 0.76 & \multicolumn{1}{c}{0.77} & 0.75 & \multicolumn{1}{c}{0.75} & 0.73 & \multicolumn{1}{c}{0.73} & 0.53 & \multicolumn{1}{c}{0.54} & 0.54 & 0.58 \\ \hline
\multicolumn{1}{l}{Auto-CoT(3-shot)} & 0.75 & \multicolumn{1}{c}{0.75} & \textbf{0.76} & \multicolumn{1}{c}{0.77} & \textbf{0.76} & \multicolumn{1}{c}{0.76} & \textbf{0.57} & \multicolumn{1}{c}{0.58} & 0.54 & 0.58 \\ \hline
\multicolumn{1}{l}{Self-Consistency} & \textbf{0.77} & \multicolumn{1}{c}{0.77} & 0.71 & \multicolumn{1}{c}{0.72} & 0.71 & \multicolumn{1}{c}{0.72} & 0.49 & \multicolumn{1}{c}{0.50} & 0.49 & 0.53 \\ \hline
\multicolumn{1}{l}{PS-CoT} & 0.74 & \multicolumn{1}{c}{0.74} & 0.72 & \multicolumn{1}{c}{0.73} & 0.73 & \multicolumn{1}{c}{0.73} & 0.48 & \multicolumn{1}{c}{0.50} & \textbf{0.56} & 0.60 \\ \hline
\multicolumn{11}{l}{\cellcolor[HTML]{EFEFEF}\textit{w/ Extra Information}} \\ \hline
\multicolumn{1}{l}{Naive w/ Choice} & 0.82 & \multicolumn{1}{c}{0.82} & 0.77 & \multicolumn{1}{c}{0.78} & 0.79 & \multicolumn{1}{c}{0.79} & 0.46 & \multicolumn{1}{c}{0.48} & 0.67 & 0.71 \\ \bottomrule
\end{tabular}}
\caption{Performance comparison on the Query-Unrelated PII Masking task (PII-single and PII-multi datasets). The best performance for each model (excluding Naive w/ Choice) is in \textbf{bold}.}
\label{tab:query-unrelated}
\end{table*}

\begin{figure*}[t]
  \includegraphics[width=2\columnwidth]{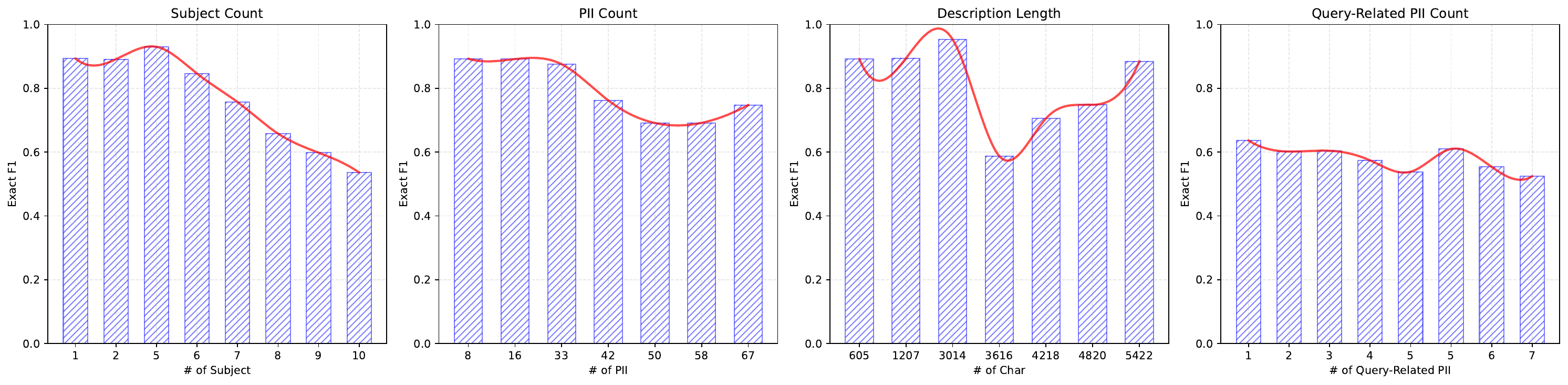}
\vspace{-2mm}
  \caption{The performance of GPT-4o is correlated with the number of subject, the number of PII, decription length, and the number of query-related PII.}
  \label{fig:factor}
\vspace{-2mm}
\end{figure*}



\begin{table}[t]
\centering
\resizebox{\columnwidth}{!}{
\begin{tabular}{lcccc}
\toprule
\multicolumn{1}{l}{} & \multicolumn{2}{c}{\textbf{Test-Hard}} & \multicolumn{2}{c}{\textbf{Test-Distract}} \\
\multicolumn{1}{l}{\multirow{-2}{*}{\textbf{Method}}} & F1 & \multicolumn{1}{c}{RougeL-F} & F1 & RougeL-F \\ \hline
\multicolumn{5}{l}{\cellcolor[HTML]{EFEFEF}\textit{Basic Method}} \\ \hline
\multicolumn{1}{l}{Naive} & 0.36 & \multicolumn{1}{c}{0.36} & 0.57 & 0.57 \\ \hline
\multicolumn{5}{l}{\cellcolor[HTML]{EFEFEF}\textit{Advanced Method}} \\ \hline
\multicolumn{1}{l}{Self-CoT} & 0.45 & \multicolumn{1}{c}{0.45} & 0.66 & 0.67 \\ \hline
\multicolumn{1}{l}{Auto-CoT(3-shot)} & 0.45 & \multicolumn{1}{c}{0.40} & 0.62 & 0.63 \\ \hline
\multicolumn{1}{l}{Self-Consistency} & \textbf{0.46} & \multicolumn{1}{c}{0.46} & 0.62 & 0.63 \\ \hline
\multicolumn{1}{l}{PS-CoT} & 0.38 & \multicolumn{1}{c}{0.38} & \textbf{0.67} & 0.67 \\ \hline
\multicolumn{5}{l}{\cellcolor[HTML]{EFEFEF}\textit{w/ Extra Information}} \\ \hline
\multicolumn{1}{l}{Naive w/ Choice} & 0.53 & \multicolumn{1}{c}{0.53} & 0.66 & 0.66 \\ \bottomrule
\end{tabular}}
\caption{Performance comparison on challenging test sets using GPT4o.}
\label{tab:hard_distract}
\vspace{-2mm}
\end{table}

\subsection{Performance on Query-Unrelated PII Masking}
We evaluate models' performance on the query-unrelated PII masking task, which requires both accurate PII detection and relevance assessment. Table~\ref{tab:query-unrelated} presents our results:

\textbf{Joint task yields improved performance.}
Models achieve higher F1 scores in this combined task compared to individual query-relevance tasks. GPT4o reaches 0.77 F1 with Self-Consistency prompting, suggesting complementary signals from the joint objective. Open-source models demonstrate comparable capabilities, with both Llama3.1 and Qwen2.5 achieving 0.76 F1 using Auto-CoT.

\subsection{Performance on PII Detection}
Results on the PII detection task (Table~\ref{tab:pii_recog}) reveal:

\textbf{Large Language Models demonstrate superior detection capabilities.}
API-based LLMs achieve strong performance, with DeepSeekV3 and GPT4o leading in Strict-F1 scores (0.903 and 0.891 on PII-Single and PII-Multi, respectively). Open-source Llama3.1 shows competitive performance, particularly in entity recognition (Ent-F1: 0.942 on PII-Multi).

\textbf{Entity type classification remains challenging.}
A consistent gap between Strict-F1 and Ent-F1 scores indicates that accurate PII type classification poses greater challenges than entity boundary detection. This disparity becomes more pronounced in the PII-Distract dataset, suggesting increased difficulty in precise PII categorization under complex scenarios.

\begin{table*}[htp]
\centering
\resizebox{2\columnwidth}{!}{
\begin{tabular}{ccccccccccccc}
\toprule
\multicolumn{1}{c}{} & \multicolumn{3}{c}{\textbf{PII-Single}} & \multicolumn{3}{c}{\textbf{PII-Multi}} & \multicolumn{3}{c}{\textbf{PII-Hard}} & \multicolumn{3}{c}{\textbf{PII-Distract}} \\
\multicolumn{1}{c}{\multirow{-2}{*}{\textbf{Baseline Models}}} & Strict-F1 & Ent-F1 & \multicolumn{1}{c}{RougeL-F} & Strict-F1 & Ent-F1 & \multicolumn{1}{c}{RougeL-F} & Strict-F1 & Ent-F1 & \multicolumn{1}{c}{RougeL-F} & Strict-F1 & Ent-F1 & RougeL-F \\ \hline
\multicolumn{13}{l}{\cellcolor[HTML]{EFEFEF}\textit{Traditional Model}} \\ \hline
\multicolumn{1}{c}{BiLSTM-CRF} & - & 0.851 & \multicolumn{1}{c}{-} & - & 0.828 & \multicolumn{1}{c}{-} & - & 0.684 & \multicolumn{1}{c}{-} & - & 0.787 & - \\ \hline
\multicolumn{13}{l}{\cellcolor[HTML]{EFEFEF}\textit{API-based Large Language Model}} \\ \hline
\multicolumn{1}{c}{GPT4o} & 0.893 & 0.914 & \multicolumn{1}{c}{0.895} & \textbf{0.891} & 0.923 & \multicolumn{1}{c}{\textbf{0.893}} & 0.817 & 0.869 & \multicolumn{1}{c}{0.819} & 0.715 & 0.868 & 0.716 \\
\multicolumn{1}{c}{Claude3.5} & 0.858 & 0.891 & \multicolumn{1}{c}{0.862} & 0.890 & 0.920 & \multicolumn{1}{c}{0.892} & 0.813 & 0.857 & \multicolumn{1}{c}{0.818} & \textbf{0.910} & \textbf{0.948} & \textbf{0.911} \\
\multicolumn{1}{c}{DeepSeekV3} & \textbf{0.903} & \textbf{0.921} & \multicolumn{1}{c}{\textbf{0.905}} & 0.884 & 0.927 & \multicolumn{1}{c}{0.886} & 0.838 & 0.893 & \multicolumn{1}{c}{0.838} & 0.658 & 0.945 & 0.658 \\ \hline
\multicolumn{13}{l}{\cellcolor[HTML]{EFEFEF}\textit{Open-source Large Language Model}} \\ \hline
\multicolumn{1}{c}{Llama3.1} & 0.881 & 0.913 & \multicolumn{1}{c}{0.883} & 0.883 & \textbf{0.942} & \multicolumn{1}{c}{0.884} & \textbf{0.840} & \textbf{0.893} & \multicolumn{1}{c}{\textbf{0.841}} & 0.834 & 0.946 & 0.835 \\
\multicolumn{1}{c}{Qwen2.5} & 0.866 & 0.908 & \multicolumn{1}{c}{0.869} & 0.853 & 0.918 & \multicolumn{1}{c}{0.855} & 0.804 & 0.876 & \multicolumn{1}{c}{0.806} & 0.647 & 0.941 & 0.649 \\ \hline
\multicolumn{13}{l}{\cellcolor[HTML]{EFEFEF}\textit{Open-source Small Language Model}} \\ \hline
\multicolumn{1}{c}{Llama3.1-SLM} & 0.748 & 0.800 & \multicolumn{1}{c}{0.752} & 0.778 & 0.869 & \multicolumn{1}{c}{0.781} & 0.718 & 0.798 & \multicolumn{1}{c}{0.722} & 0.551 & 0.876 & 0.552 \\
\multicolumn{1}{c}{Qwen2.5-SLM} & 0.787 & 0.846 & \multicolumn{1}{c}{0.792} & 0.451 & 0.806 & \multicolumn{1}{c}{0.453} & 0.591 & 0.810 & \multicolumn{1}{c}{0.594} & 0.454 & 0.815 & 0.456 \\ \bottomrule
\end{tabular}}
\caption{Performance of baseline models under the PII Detection task. Results in \textbf{bold} indicate the best performance for each dataset and metric category.}
\label{tab:pii_recog}
\vspace{-2mm}
\end{table*}

\begin{table*}[htp]
\centering
\resizebox{2\columnwidth}{!}{
\begin{tabular}{lcccccccccc}
\toprule
\multicolumn{1}{l}{} & \multicolumn{2}{c}{\textbf{GPT4o}} & \multicolumn{2}{c}{\textbf{Llama3.1}} & \multicolumn{2}{c}{\textbf{Qwen2.5}} & \multicolumn{2}{c}{\textbf{Llama3.1-SLM}} & \multicolumn{2}{c}{\textbf{Qwen2.5-SLM}} \\
\multicolumn{1}{l}{\multirow{-2}{*}{\textbf{Method}}} & F1 & \multicolumn{1}{c}{RougeL-F} & F1 & \multicolumn{1}{c}{RougeL-F} & F1 & \multicolumn{1}{c}{RougeL-F} & F1 & \multicolumn{1}{c}{RougeL-F} & F1 & RougeL-F \\ \hline
\multicolumn{11}{l}{\cellcolor[HTML]{EFEFEF}\textit{Basic Method}} \\ \hline
\multicolumn{1}{l}{Naive} & 0.63 & \multicolumn{1}{c}{0.63} & 0.63 & \multicolumn{1}{c}{0.63} & 0.62 & \multicolumn{1}{c}{0.62} & 0.33 & \multicolumn{1}{c}{0.33} & 0.41 & 0.41 \\ \hline
\multicolumn{11}{l}{\cellcolor[HTML]{EFEFEF}\textit{Advanced Method}} \\ \hline
\multicolumn{1}{l}{Self-CoT} & 0.71 & \multicolumn{1}{c}{0.72} & 0.69 & \multicolumn{1}{c}{0.69} & 0.67 & \multicolumn{1}{c}{0.68} & 0.39 & \multicolumn{1}{c}{0.39} & 0.40 & 0.41 \\ \hline
\multicolumn{1}{l}{Auto-CoT(3-shot)} & 0.66 & \multicolumn{1}{c}{0.66} & \textbf{0.70} & \multicolumn{1}{c}{0.72} & \textbf{0.71} & \multicolumn{1}{c}{0.72} & \textbf{0.43} & \multicolumn{1}{c}{0.44} & 0.37 & 0.38 \\ \hline
\multicolumn{1}{l}{Self-Consistency} & \textbf{0.72} & \multicolumn{1}{c}{0.72} & 0.63 & \multicolumn{1}{c}{0.64} & 0.65 & \multicolumn{1}{c}{0.65} & 0.31 & \multicolumn{1}{c}{0.32} & 0.32 & 0.33 \\ \hline
\multicolumn{1}{l}{PS-CoT} & 0.65 & \multicolumn{1}{c}{0.65} & 0.65 & \multicolumn{1}{c}{0.66} & 0.67 & \multicolumn{1}{c}{0.67} & 0.35 & \multicolumn{1}{c}{0.36} & \textbf{0.45} & 0.46 \\ \hline
\multicolumn{11}{l}{\cellcolor[HTML]{EFEFEF}\textit{w/ Extra Information}} \\ \hline
\multicolumn{1}{l}{Naive w/ Choice} & 0.84 & \multicolumn{1}{c}{0.84} & 0.76 & \multicolumn{1}{c}{0.76} & 0.83 & \multicolumn{1}{c}{0.83} & 0.52 & \multicolumn{1}{c}{0.52} & 0.77 & 0.77 \\ \bottomrule
\end{tabular}}
\caption{Performance comparison on the Query-Related PII Detection task (PII-single dataset).}
\label{tab:single}
\vspace{-5mm}
\end{table*}

\subsection{Performance on Query-Related PII Detection}
Table \ref{tab:single} presents the results on PII-single dataset across different model scales and prompting strategies. 

\textbf{Limited Performance of Current LLMs.} Advanced LLMs exhibit limited performance, with GPT4o achieving only 0.627 F1 score with naive prompting—substantially below human performance (0.951 F1). Open-source alternatives show competitive performance, with Qwen2.5 reaching 0.615 F1.

\textbf{Impact of Advanced Prompting.} Chain-of-thought approaches generally improve performance, with Self-Consistency and Auto-CoT proving most effective (0.716 F1 for GPT4o with Self-Consistency; 0.710 F1 for Qwen2.5 with Auto-CoT). However, these benefits are highly dependent on model scale—smaller models often show degraded performance with complex prompting strategies.

\textbf{Effectiveness of Entity Candidates.} Providing candidate PII entities (Naive w/ Choice) substantially improves performance across all models (e.g., GPT4o improves from 0.627 to 0.842 F1). However, practical applicability is limited as candidate entities are rarely available in real-world scenarios.

\subsection{Privacy-Utility Tradeoff Analysis}

We evaluated the tradeoff between privacy protection and query utility across different PII masking strategies

\textbf{Experimental Setup.} We selected 200 unique user descriptions from PII-Bench and applied three masking methods: (1) \textbf{No Mask}: Original text with all PII preserved; (2) \textbf{All PII Mask}: All detected PII entities replaced with their corresponding tags; (3) \textbf{Query-unrelated PII Mask}: Only query-irrelevant PII entities masked. 
For each method, we generated responses using GPT4o.

\textbf{Metrics.} The privacy-utility tradeoff is evaluated through two metrics: \textbf{Privacy Score ($P$)}: Quantifies the proportion of PII successfully protected in the processed text. \textbf{Utility Score ($U$)}: Combines semantic similarity and LLM evaluation for quality assessment between mask and unmasked responses. Detailed computation procedures are provided in Appendix \ref{sec:additional_metrics}.

\textbf{Results.} Table \ref{tab:privacy-utility} shows that Query-unrelated PII Mask achieves optimal balance between privacy protection ($P=0.83$) and utility preservation ($U=0.89$), outperforming other strategies in balanced score ($B=0.86$).

\begin{table}[t]
\centering
\small
\begin{tabular}{lccc}
\toprule
\textbf{Method} & \textbf{P} & \textbf{U} & \textbf{B} \\
\hline
No Mask & 0.00 & 1.00 & 0.50 \\
All PII Mask & 1.00 & 0.52 & 0.76 \\
Query-unrelated PII Mask & 0.83 & 0.89 & 0.86 \\
\bottomrule
\end{tabular}
\caption{Privacy-utility tradeoff across different PII masking strategies. Balanced scores ($B$) combine both metrics with equal weights.}
\label{tab:privacy-utility}
\end{table}

\subsection{Real-World Dataset Validation}
\label{sec:pii_real_validation}
To validate the reliability of synthetic data, we constructed PII-Real, a dataset comprising 100 instances derived from publicly available profiles of 20 AI researchers with manually annotated PII entities and human-written queries. 
Evaluation on PII-Real demonstrates consistent alignment with synthetic data. Comprehensive experimental results and validation analysis are provided in Appendix~\ref{sec:pii_real}.

\subsection{In-depth Performance Analysis}


\textbf{Factors Influencing Performance.}
Figure~\ref{fig:factor} reveals several critical factors affecting model accuracy: performance degrades sharply beyond 5 subjects (F1 drops from 0.85 to 0.52), 33 PII entities, or 3000 characters in text length. Query-Related entity count shows modest impact, with gradual decline from 1 to 7 entities.

\textbf{Performance on Challenging Scenarios.}
Results on specialized test sets (Table~\ref{tab:hard_distract}) reveal significant performance degradation in complex scenarios. On Test-Hard, featuring high PII density and long texts, even the best-performing Self-Consistency approach achieves only 0.463 F1. Test-Distract's multi-subject scenarios pose similar challenges.

\section{Conclusion}
This paper introduces PII-Bench, a comprehensive evaluation framework comprising 2,842 test samples across 7 PII types with 55 fine-grained subcategories, and proposes a query-unrelated PII masking strategy to balance privacy protection with LLM utility. Our empirical evaluation reveals that while advanced LLMs demonstrate strong performance in basic PII detection, they exhibit substantial limitations in query-relevance assessment. Small-scale models show considerably larger performance gaps across all evaluation tasks. These findings establish foundational benchmarks and exposes critical challenges in privacy-aware PII handling.

\section*{Limitations}
Despite PII-Bench's contributions to privacy protection evaluation, several limitations merit acknowledgment. While the current dataset encompasses common privacy scenarios, it requires expansion into specialized domains such as medical records and financial transactions. Our automated synthesis methodology mitigates this limitation by enabling flexible dataset expansion across domains, languages, and cultural contexts, supporting continuous refinement of PII categories to meet evolving application requirements.
The evaluation framework primarily assesses the accuracy of PII entity detection and query relevance determination, but lacks systematic evaluation of models' reasoning processes. Specifically, it does not fully capture how models interpret queries, derive information requirements, and make relevance judgments about sensitive information. This gap in assessment methodology limits our understanding of models' reasoning capabilities in real-world privacy protection scenarios.

\section*{Ethical Concerns}
Throughout the development and implementation of PII-Bench, ethical considerations have remained our paramount priority. To ensure the evaluation dataset itself does not compromise privacy, we have implemented rigorous data synthesis and review protocols, with all sample data undergoing multiple rounds of scrutiny by professional security teams to guarantee the absence of real personal information. During the data generation process, we have carefully engineered our algorithms to ensure equitable representation across different demographic groups, establishing comprehensive human review mechanisms to verify that generated data remains free from bias and discriminatory content. 
\bibliography{latex/custom}

\appendix

\section{Details about PII}
\label{sec:pii}
\subsection{PII Definition}
\label{sec:def}
In this section, we follow previous work by categorizing Personally Identifiable Information (PII) into the following two categories(\citealp{elliot2016anonymisation},\citealp{domingo2022database},\citealp{papadopoulou2022neural}):
\begin{itemize}
    \item \textit{Direct identifiers}: Information that can uniquely identify an individual within a dataset(e.g. name, social security number, email address, etc). 
    \item \textit{Quasi identifiers}: Information that cannot uniquely identify an individual on their own but can do so when combined with other quasi-identifiers(e.g. age, gender, occupation, etc.
\end{itemize}

Because of their high sensitivity or the potential to indirectly identify an individual, both direct and quasi-identifiers are governed by strict legal and privacy standards to ensure personal privacy.

\subsection{PII Types}
\label{sec:type}
Unlike \citeposs{papadopoulou2022neural} classification, we exclude the MISC category due to its ambiguous definition and unclear boundaries. Our taxonomy comprises seven categories:

\textbf{PER}: Refers to individuals' names, including full names, aliases, and social media usernames.

\textbf{CODE}: Encompasses identifying numbers and codes like social security numbers, phone numbers, passport numbers, email addresses, etc.

\textbf{LOC}: Covers geographical locations such as home or work addresses, cities, countries, etc.

\textbf{ORG}: Pertains to the names of entities like companies, schools, public institutions, etc.

\textbf{DEM}: Represents demographic information including age, gender, nationality, occupation, education level, etc.

\textbf{DATETIME}: Indicates specific dates, times, or durations, such as birthdates, appointment times, etc.

\textbf{QUANTITY}: Refers to significant numerical data like monthly income, expenditures, loan amount, credit score, etc.

\subsection{Statistics of PII Types}
\begin{figure}[t]
  \includegraphics[width=\columnwidth]{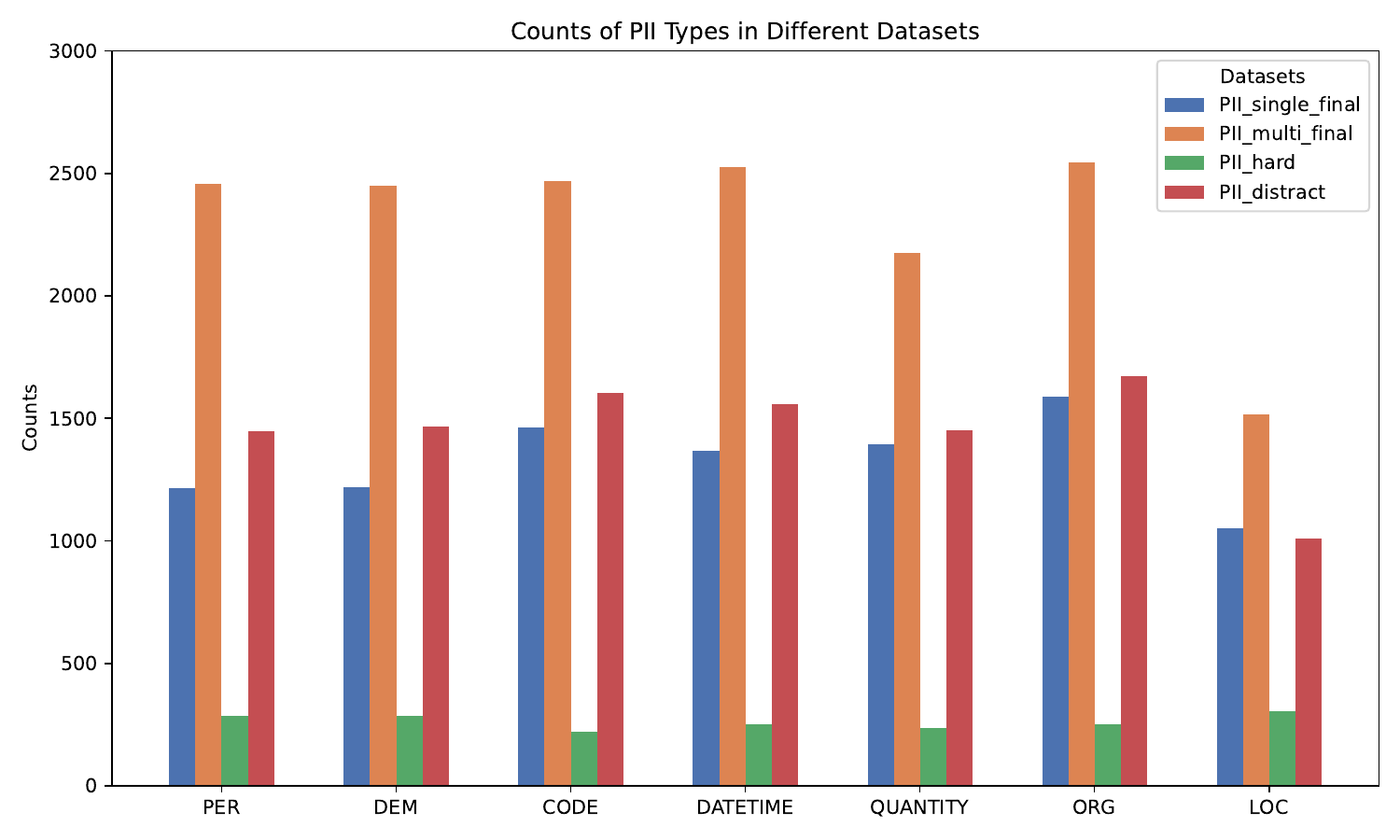}
  \caption{Distribution of PII types across different datasets in PII-Bench.}
  \label{fig:pii_dist}
\end{figure}

Figure~\ref{fig:pii_dist} and Table~\ref{tab:pii_stats} present the distribution of PII types across our datasets: PII-single (1,214 samples), PII-multi (1,228 samples), PII-hard (200 samples), and PII-distract (200 samples). 

\begin{itemize}
\item \textbf{Type Frequencies}: Organization (ORG) and Code-based identifiers (CODE) constitute significant portions across all datasets, with 17.09\% and 15.74\% in PII-single, and 13.47\% and 15.31\% in PII-multi, respectively. This distribution reflects the prevalence of institutional affiliations and digital identifiers in real-world scenarios.

\item \textbf{Dataset Composition}: PII-multi contains 16,136 PII entities across all categories, maintaining balanced proportions ranging from 13.47\% to 15.77\% for most types. PII-single follows a similar pattern with 9,303 entities, demonstrating consistent coverage across different PII categories.

\item \textbf{Specialized Test Sets}: PII-distract, despite comprising only 200 samples, contains 10,211 PII entities due to its multi-description design. PII-hard maintains balanced type coverage with 1,834 entities, with proportions varying from 12.10\% to 16.58\%.
\end{itemize}

\section{PII Entity Generation Methods}
\label{sec:ent_gen}
The generation of PII entities requires careful consideration of both structural constraints and semantic plausibility. We employ two complementary approaches for entity generation: rule-based generation for structured PII types and language model-based generation for context-dependent information.
\subsection{Rule-based Generation}
For PII types with well-defined formats or enumerable value sets, we implement deterministic generation methods. These methods encompass both custom rule-based algorithms and the Faker library's standardized functions. The rule-based approach is particularly effective for:

\begin{enumerate}
    \item Identification Numbers: Generating valid formats for social security numbers, passport numbers, and employee IDs while maintaining regional compliance.
    \item Contact Information: Creating syntactically correct email addresses, phone numbers, and IP addresses.
    \item Financial Data: Producing properly formatted credit card numbers, bank account numbers, and other numerical identifiers with appropriate check digits.
    \item Temporal Information: Generating dates, times, and durations within reasonable ranges and formats.
\end{enumerate}

\begin{table}[htp]
\centering
\resizebox{\columnwidth}{!}{
\begin{tabular}{ccccccccc}
\toprule
\multirow{2}{*}{\textbf{Type}} &
  \multicolumn{2}{c}{\textbf{PII\_single}} &
  \multicolumn{2}{c}{\textbf{PII\_multi}} &
  \multicolumn{2}{c}{\textbf{PII\_hard}} &
  \multicolumn{2}{c}{\textbf{PII\_distract}} \\ \cline{2-9} 
                  & \#    & \%    & \#     & \%    & \#    & \%    & \#     & \%    \\ \hline
\textbf{PER}      & 1,214 & 13.05 & 2,456  & 15.22 & 286   & 15.59 & 1,449  & 14.19 \\
\textbf{DEM}      & 1,220 & 13.12 & 2,450  & 15.18 & 286   & 15.59 & 1,467  & 14.37 \\
\textbf{CODE}     & 1,464 & 15.74 & 2,470  & 15.31 & 222   & 12.10 & 1,605  & 15.72 \\
\textbf{ORG}      & 1,590 & 17.09 & 2,544  & 13.47 & 251   & 13.69 & 1,673  & 16.38 \\
\textbf{LOC}      & 1,053 & 11.32 & 1,516  & 15.77 & 304   & 16.58 & 1,008  & 9.87  \\
\textbf{DATETIME} & 1,368 & 14.70 & 2,526  & 15.65 & 251   & 13.69 & 1,559  & 15.27 \\
\textbf{QUANTITY} & 1,394 & 14.98 & 2,174  & 9.40  & 234   & 12.76 & 1,450  & 14.20 \\ \hline
\textbf{Total}    & 9,303 & 100   & 16,136 & 100   & 1,834 & 100   & 10,211 & 100   \\ \toprule
\end{tabular}}
\caption{Detailed statistics of PII types across datasets. For each dataset, we report both the absolute count (\#) and relative percentage (\%) of each PII type.}
\label{tab:pii_stats}
\end{table}

\subsection{Language Model-based Generation}
For PII types requiring contextual understanding and real-world knowledge, we leverage large language models through carefully designed prompts. This approach is essential for generating:

\begin{enumerate}
    \item Location Information: Coherent and geographically accurate addresses, landmarks, and regional descriptions.
    \item Organizational Entities: Plausible names for educational institutions, companies, and other organizations that reflect real-world naming conventions.
    \item Demographic Attributes: Culturally appropriate and consistent demographic information, including ethnicity, nationality, and educational background.
\end{enumerate}

\subsection{Entity Categories and Generation Methods}
Table~\ref{tab:ent_gen} presents a comprehensive mapping of PII types to their respective generation methods. The table systematically categorizes 55 distinct PII entities across seven main categories: Personal Identifiers (PER), Codes and Numbers (CODE), Location Information (LOC), Organizational Affiliations (ORG), Demographic Information (DEM), Temporal Data (DATETIME), and Quantitative Values (QUANTITY).

\begin{table*}[htp]
\centering
\small
\begin{tabular}{llll}
\toprule
\textbf{PII Type} & \textbf{Entity Category} & \textbf{Generation Approach} & \textbf{Format Constraints} \\
\midrule
\multirow{3}{*}{\textbf{PER}} 
    & Full Name & Rule-based & [First Name] [Last Name] \\
    & Social Media Handle & Rule-based & [@][a-zA-Z0-9]{5,15} \\
    & Nickname & LLM-based & - \\
\midrule
\multirow{12}{*}{\textbf{CODE}} 
    & Social Security Number & Rule-based & XXX-XX-XXXX \\
    & Driver's License & Rule-based & [A-Z][0-9]{8} \\
    & Bank Account & Rule-based & [0-9]{10,12} \\
    & Credit Card & Rule-based & [0-9]{16} \\
    & Phone Number & Rule-based & +[0-9]{1,3}-[0-9]{10} \\
    & IP Address & Rule-based & IPv4/IPv6 format \\
    & Email Address & Rule-based & [user]@[domain].[tld] \\
    & Password Hash & Rule-based & SHA-256 \\
    & Passport Number & Rule-based & [A-Z][0-9]{8} \\
    & Tax ID & Rule-based & [0-9]{9} \\
    & Employee ID & Rule-based & [A-Z]{2}[0-9]{6} \\
    & Student ID & Rule-based & [0-9]{8} \\
\midrule
\multirow{3}{*}{\textbf{LOC}} 
    & Street Address & LLM-based & - \\
    & City/Region & LLM-based & - \\
    & Landmark & LLM-based & - \\
\midrule
\multirow{5}{*}{\textbf{ORG}} 
    & Company Name & LLM-based & - \\
    & Educational Institution & LLM-based & - \\
    & Government Agency & LLM-based & - \\
    & NGO & LLM-based & - \\
    & Healthcare Facility & LLM-based & - \\
\midrule
\multirow{17}{*}{\textbf{DEM}} 
    & Occupation & Rule-based & Predefined list \\
    & Age & Rule-based & [0-9]{1,3} \\
    & Gender & Rule-based & Binary/Non-binary \\
    & Height & Rule-based & [0-9]{3}cm/[0-9]'[0-9]" \\
    & Weight & Rule-based & [0-9]{2,3}kg/lbs \\
    & Blood Type & Rule-based & A/B/O[+-] \\
    & Sexual Orientation & Rule-based & Predefined list \\
    & Nationality & LLM-based & - \\
    & Ethnicity & LLM-based & - \\
    & Race & LLM-based & - \\
    & Religious Belief & LLM-based & - \\
    & Political Affiliation & LLM-based & - \\
    & Education Level & LLM-based & - \\
    & Academic Degree & LLM-based & - \\
    & Physical Features & LLM-based & - \\
    & Medical Condition & LLM-based & - \\
    & Disability Status & LLM-based & - \\
\midrule
\multirow{3}{*}{\textbf{DATETIME}} 
    & Date & Rule-based & YYYY-MM-DD \\
    & Time & Rule-based & HH:MM:SS \\
    & Duration & Rule-based & [0-9]+[dhms] \\
\midrule
\multirow{12}{*}{\textbf{QUANTITY}} 
    & Monthly Income & Rule-based & [Currency][0-9]+ \\
    & Monthly Expenses & Rule-based & [Currency][0-9]+ \\
    & Account Balance & Rule-based & [Currency][0-9]+ \\
    & Loan Amount & Rule-based & [Currency][0-9]+ \\
    & Annual Bonus & Rule-based & [Currency][0-9]+ \\
    & Credit Limit & Rule-based & [Currency][0-9]+ \\
    & Social Security Payment & Rule-based & [Currency][0-9]+ \\
    & Tax Payment & Rule-based & [Currency][0-9]+ \\
    & Debt Ratio & Rule-based & [0-9]{1,2}.[0-9]{2}\% \\
    & Investment Return & Rule-based & [0-9]{1,2}.[0-9]{2}\% \\
    & ROI & Rule-based & [0-9]{1,2}.[0-9]{2}\% \\
    & Credit Score & Rule-based & [300-850] \\
\bottomrule
\end{tabular}
\caption{Comprehensive categorization of PII entities and their generation methods. Rule-based generation follows specific format constraints, while LLM-based generation produces contextually appropriate content without rigid formatting requirements.}
\label{tab:ent_gen}
\end{table*}

\section{Consistency Optimization Details}
\label{sec:consistency_opt}
The consistency optimization process is critical for ensuring that randomly sampled PII entities form logically coherent and realistic profiles. We employ GPT-4-0806 with carefully designed prompts to identify and resolve conflicts while preserving the diversity and coverage of PII types.

\subsection{Single-Subject Consistency Optimization}
For single-subject descriptions, the optimization process addresses intra-subject conflicts arising from incompatible entity combinations. Common conflict patterns include:

\begin{itemize}
    \item \textbf{Temporal inconsistencies}: Age incompatible with work experience duration, education level, or career stage (e.g., a 23-year-old with 15 years of work experience).
    \item \textbf{Professional mismatches}: Occupation inconsistent with education background or salary range (e.g., a high school graduate working as a licensed physician).
    \item \textbf{Geographic contradictions}: Work location distant from residential address without supporting evidence (e.g., daily commute spanning different continents).
    \item \textbf{Financial implausibility}: Income, expenses, and savings that violate basic economic constraints (e.g., monthly expenses exceeding monthly income by orders of magnitude).
\end{itemize}

The optimization prompt (Figure~\ref{fig:consistency-single}) instructs the model to: (1) identify logical conflicts among sampled entities; (2) modify only the entity values while preserving their PII type and category labels; (3) maximize entity retention to maintain dataset richness. For instance, given a conflict between ``Age: 22 years'' and ``Work Experience: 15 years as Senior Engineer'', the optimization adjusts the work experience to ``2 years as Junior Engineer'' rather than removing the entity entirely, thereby preserving both the PII type (DEM) and the entity category (Work Experience).

\subsection{Multi-Subject Consistency Optimization}
Multi-subject optimization extends the single-subject process by introducing relationship-aware constraints. Consistency rules are enforced based on three relationship categories:

\begin{itemize}
    \item \textbf{Intersection relationships} (e.g., colleagues, classmates): Subjects must share critical attributes such as organization name, work location for colleagues, or educational institution for classmates. The optimization verifies attribute alignment and adjusts inconsistent entities accordingly.
    
    \item \textbf{Hierarchical relationships} (e.g., parent-child, supervisor-subordinate): Structural constraints are enforced such as age differences (parent at least 18-20 years older than child), authority levels (supervisor holding higher position than subordinate), and derived attributes (children inheriting nationality or ethnicity from parents when culturally appropriate).
    
    \item \textbf{Non-intersection relationships}: Subjects maintain independent entity sets with no enforced dependencies, though internal consistency within each subject's profile is still verified.
\end{itemize}

The multi-subject prompt (Figure~\ref{fig:consistency-multi}) provides relationship context and requires the model to output separate, coordinated entity sets for each subject. This relationship-aware approach ensures that multi-subject descriptions reflect realistic interpersonal dynamics and maintain narrative coherence across profiles.

\section{Query Scenario Design Details}
\label{sec:scenario_design}
The scenario design phase bridges the gap between abstract entity selection and natural user queries by grounding queries in realistic application contexts. This process is essential for evaluating whether privacy protection systems can identify query-relevant PII in authentic usage scenarios.

\subsection{Domain-Specific Scenario Generation}
Given a set of selected entities $\mathcal{E}_q$, we employ a structured prompting strategy to generate diverse scenario contexts. The prompt instructs the LLM to:

\begin{enumerate}
    \item \textbf{Analyze entity semantics}: Determine the thematic domain implied by the selected entities (e.g., education and work experience suggest career-related scenarios; medical conditions and medications suggest healthcare scenarios).
    
    \item \textbf{Generate scenario candidates}: Propose multiple scenario types that naturally require all selected entities, ensuring thematic diversity across the dataset. Example scenarios include career planning, medical consultation, financial advisory, legal counseling, academic mentoring, and housing applications.
    
    \item \textbf{Ensure entity necessity}: Verify that each selected entity is meaningfully incorporated into the scenario rather than superficially included, so that query-relevant PII labels reflect genuine information dependencies.
\end{enumerate}

The scenario generation prompt is provided in Figure~\ref{fig:scenario-prompt}. By requiring coverage of all selected entities while maintaining scenario diversity, this approach produces queries that realistically reflect specialized user information needs across domains.

\subsection{Scenario Validation and Filtering}
Generated scenarios undergo validation to ensure quality and diversity:

\begin{itemize}
    \item \textbf{Completeness check}: Verify that all entities in $\mathcal{E}_q$ are semantically incorporated into the scenario context.
    \item \textbf{Realism assessment}: Confirm that the scenario represents plausible real-world interactions with AI systems.
    \item \textbf{Diversity enforcement}: Track scenario distribution and reject over-represented scenario types to maintain balanced domain coverage.
\end{itemize}

This rigorous scenario design process ensures that PII-Bench queries reflect authentic information needs across diverse domains, providing a realistic testbed for evaluating query-aware privacy protection systems.

\section{Human Evaluation Details}
\label{sec:human_eval}
The human evaluation of PII-Bench was conducted with 25 graduate students specializing in data security and privacy protection. All evaluators were pursuing their Master's or Ph.D. degrees with at least two years of research experience in privacy-preserving machine learning or data protection systems. The evaluation process consisted of three phases: preparation, evaluation, and validation.

During the preparation phase, participants attended a 4-hour training session covering PII taxonomy, recognition guidelines, and query-related detection criteria. The session included hands-on practice with representative cases from each dataset component. Participants then completed a qualification test featuring 20 diverse instances, requiring 90\% agreement with expert assessments to proceed to the formal evaluation.

During the evaluation phase, participants used our specialized platform designed for systematic PII assessment. To maintain consistent performance, we limited evaluation sessions to two hours and distributed instances across a two-week period. The platform automatically tracked assessment time and accuracy metrics while enforcing our evaluation protocol: participants first performed PII detection by marking entity spans, linking them to subjects, and assigning PII types, before proceeding to query-related detection.

Our validation process incorporated both automated and manual checks to ensure assessment quality. The platform automatically verified assessment completeness and format consistency. Cases with substantial disagreement (Fleiss' kappa < 0.6) underwent expert review by two authors with extensive experience in privacy-preserving systems. Evaluators received detailed feedback on their performance and participated in discussion sessions to resolve systematic discrepancies.

Compensation was structured to encourage both accuracy and efficiency, with a base rate of \$30 per hour and performance bonuses based on agreement with other evaluators.

\section{Experiment Details}
\subsection{Evaluation Metrics}
\label{sec:metrics}
The evaluation framework employs distinct metrics for PII detection and query-related detection tasks.

For PII detection, let $\mathcal{P} = \{p_1, ..., p_m\}$ denote the predicted subject set and $\mathcal{G} = \{g_1, ..., g_n\}$ denote the ground truth subject set. Each subject $p_i$ or $g_j$ contains a set of entity-type pairs $\{(e, t)\}$, where $e$ represents the entity span and $t$ represents its PII type.

For each subject pair $(p_i, g_j)$, we compute three types of evaluation metrics:

\noindent\textbf{1. Strict Matching:} Both entity spans and their types must match exactly:
\begin{equation}
    P_{strict}(p_i, g_j) = \frac{|E_{p_i} \cap E_{g_j}|}{|E_{p_i}|}
\end{equation}
\begin{equation}
    R_{strict}(p_i, g_j) = \frac{|E_{p_i} \cap E_{g_j}|}{|E_{g_j}|}
\end{equation}
\begin{equation}
    F1_{strict}(p_i, g_j) = \frac{2 \cdot P_{strict}(p_i, g_j) \cdot R_{strict}(p_i, g_j)}{P_{strict}(p_i, g_j) + R_{strict}(p_i, g_j)}
\end{equation}
where $E_{p_i}$ and $E_{g_j}$ are the sets of entity-type pairs.

\noindent\textbf{2. Entity-only Matching:} Only entity spans need to match:
\begin{equation}
    P_{ent}(p_i, g_j) = \frac{|S_{p_i} \cap S_{g_j}|}{|S_{p_i}|}
\end{equation}
\begin{equation}
    R_{ent}(p_i, g_j) = \frac{|S_{p_i} \cap S_{g_j}|}{|S_{g_j}|}
\end{equation}
\begin{equation}
    F1_{ent}(p_i, g_j) = \frac{2 \cdot P_{ent}(p_i, g_j) \cdot R_{ent}(p_i, g_j)}{P_{ent}(p_i, g_j) + R_{ent}(p_i, g_j)}
\end{equation}
where $S_{p_i}$ and $S_{g_j}$ are the sets of entity spans.

The optimal subject matching $M^*$ is determined by maximizing the strict F1 score:
\begin{equation}
    M^* = \max_{M \in \mathcal{M}} \sum_{(p_i, g_j) \in M} F1_{strict}(p_i, g_j)
\end{equation}
where $\mathcal{M}$ denotes all possible one-to-one mappings between predicted and ground truth subjects.

The final recognition scores are computed over the optimal matching pairs:
\begin{equation}
    P_{strict} = \frac{1}{|\mathcal{P}|} \sum_{(p_i, g_j) \in M^*} P_{strict}(p_i, g_j)
\end{equation}
\begin{equation}
    R_{strict} = \frac{1}{|\mathcal{G}|} \sum_{(p_i, g_j) \in M^*} R_{strict}(p_i, g_j)
\end{equation}
\begin{equation}
    F1_{strict} = \frac{1}{max(\mathcal{P},\mathcal{G})} \sum_{(p_i, g_j) \in M^*} F1_{strict}(p_i, g_j)
\end{equation}

$P_{span}$, $R_{span}$, and $F1_{span}$ are computed analogously.

For query-related detection, given a predicted entity set $\mathcal{E}_p$ and ground truth set $\mathcal{E}_g$, we compute:
\begin{equation}
    P_{query} = \frac{|\mathcal{E}_p \cap \mathcal{E}_g|}{|\mathcal{E}_p|}
\end{equation}
\begin{equation}
    R_{query} = \frac{|\mathcal{E}_p \cap \mathcal{E}_g|}{|\mathcal{E}_g|}
\end{equation}
\begin{equation}
    F1_{query} = \frac{2 \cdot P_{query} \cdot R_{query}}{P_{query} + R_{query}}
\end{equation}

For both PII detection and query-related detection tasks, we additionally employ Rouge-L based fuzzy matching to handle partial matches between entity spans. Instead of using exact set intersection, the Rouge-L score is used to measure textual similarity between entities:
\begin{equation}
    P_{fuzzy} = \frac{1}{|\mathcal{E}_p|} \sum_{e_p \in \mathcal{E}_p} \max_{e_g \in \mathcal{E}_g} Rouge\text{-}L(e_p, e_g)
\end{equation}
\begin{equation}
    R_{fuzzy} = \frac{1}{|\mathcal{E}_g|} \sum_{e_g \in \mathcal{E}_g} \max_{e_p \in \mathcal{E}_p} Rouge\text{-}L(e_p, e_g)
\end{equation}
\begin{equation}
    F1_{fuzzy} = \frac{2 \cdot P_{fuzzy} \cdot R_{fuzzy}}{P_{fuzzy} + R_{fuzzy}}
\end{equation}
where $Rouge\text{-}L(e_p, e_g)$ computes the longest common subsequence-based F-score between predicted entity $e_p$ and ground truth entity $e_g$.

\subsection{Privacy-Utility Tradeoff Metrics}
\label{sec:additional_metrics}
We develop a comprehensive evaluation framework to quantify the tradeoff between privacy protection and query utility across different PII masking strategies.

\subsubsection{Privacy Protection Metric}

Let $\mathcal{E} = \{e_1, ..., e_n\}$ denote the set of PII entities in the original text $T_o$, and $T_m$ represent the masked text. The privacy score $P$ measures the proportion of PII entities successfully protected:

\begin{equation}
    P = 1 - \frac{\sum_{e \in \mathcal{E}} C(e, T_m)}{\sum_{e \in \mathcal{E}} C(e, T_o)}
\end{equation}

where $C(e, T)$ counts occurrences of entity $e$ in text $T$. A score of $P = 1$ indicates complete protection, while $P = 0$ indicates no protection.

\subsubsection{Utility Preservation Metrics}

To measure utility preservation, we employ two complementary approaches:

\paragraph{Semantic Similarity} We compute embedding-based similarity between responses generated from masked prompts ($R_m$) and unmasked prompts ($R_o$):

\begin{equation}
    U_s = \cos(\mathbf{v}_{R_m}, \mathbf{v}_{R_o})
\end{equation}

where $\mathbf{v}_{R}$ is the text embedding generated by BGE-M3 \cite{multi2024m3}, and $\cos(\cdot,\cdot)$ computes cosine similarity. This metric captures semantic preservation independent of exact wording.

\paragraph{LLM-as-Judge Evaluation} We employ Claude-3.7-Sonnet to assess response quality through direct comparison:

\begin{equation}
    U_l = Judge(R_m^1, R_m^2, R_o)
\end{equation}

where $R_m^1$ and $R_m^2$ represent responses from two different masking strategies, and $R_o$ is the reference response from unmasked text. The judge assigns numerical ratings $r \in [1,10]$ to each masked response based on how well it preserves the query intent compared to the reference. To mitigate position bias, we randomly alternate the presentation order of responses.

\subsubsection{Balanced Metric}

To quantify the overall effectiveness of masking strategies, we compute a balanced score $B$ that combines privacy protection and utility preservation:

\begin{equation}
    B = \alpha P + (1-\alpha) U
\end{equation}

where $\alpha \in [0,1]$ is a weighting parameter that determines the relative importance of privacy versus utility. In our experiments, we use $\alpha = 0.5$ to assign equal importance to both aspects.

The complete prompt template for LLM-as-Judge evaluation is provided in Figure \ref{fig:llm-judge-prompt}.

\subsection{Additional Experiments with Smaller Language Models}
\label{sec:small-models}
We have conducted a comprehensive evaluation on smaller, deployment-ready models (0.5B-3B parameters) to assess their viability for on-device PII protection.

\begin{table*}[htp]
\centering
\resizebox{2\columnwidth}{!}{
\begin{tabular}{lcccccccccccc}
\toprule
& \multicolumn{3}{c}{\textbf{PII-Single}} & \multicolumn{3}{c}{\textbf{PII-Multi}} & \multicolumn{3}{c}{\textbf{PII-Hard}} & \multicolumn{3}{c}{\textbf{PII-Distract}} \\
\cmidrule(lr){2-4} \cmidrule(lr){5-7} \cmidrule(lr){8-10} \cmidrule(lr){11-13}
\textbf{Baseline Models} & \textbf{Strict-F1} & \textbf{Ent-F1} & \textbf{RougeL-F} & \textbf{Strict-F1} & \textbf{Ent-F1} & \textbf{RougeL-F} & \textbf{Strict-F1} & \textbf{Ent-F1} & \textbf{RougeL-F} & \textbf{Strict-F1} & \textbf{Ent-F1} & \textbf{RougeL-F} \\
\midrule
Qwen2.5-0.5B & 0.002 & 0.0029 & 0.002 & 0.0013 & 0.0042 & 0.0013 & 0.0034 & 0.0042 & 0.0034 & 0.009 & 0.0207 & 0.009 \\
Qwen2.5-1.5B & 0.1846 & 0.2069 & 0.1865 & 0.1057 & 0.1794 & 0.1071 & 0.1474 & 0.1976 & 0.1502 & 0.0728 & 0.1549 & 0.0733 \\
Qwen2.5-3B & 0.5929 & 0.6289 & 0.5962 & 0.6189 & 0.7067 & 0.6212 & 0.5202 & 0.5921 & 0.5237 & 0.3717 & 0.6293 & 0.3731 \\
\bottomrule
\end{tabular}}
\caption{Performance of baseline models under the PII Detection task.}
\label{tab:slm-1}
\end{table*}

\begin{table}[htp]
\centering
\resizebox{1\columnwidth}{!}{
\begin{tabular}{lcccccc}
\toprule
& \multicolumn{2}{c}{\textbf{Qwen2.5-0.5B}} & \multicolumn{2}{c}{\textbf{Qwen2.5-1.5B}} & \multicolumn{2}{c}{\textbf{Qwen2.5-3B}} \\
\cmidrule(lr){2-3} \cmidrule(lr){4-5} \cmidrule(lr){6-7}
\textbf{Method} & \textbf{F1} & \textbf{RougeL-F} & \textbf{F1} & \textbf{RougeL-F} & \textbf{F1} & \textbf{RougeL-F} \\
\midrule
\multicolumn{7}{l}{\cellcolor[HTML]{EFEFEF}\textit{Basic Method}} \\
\midrule
Naive & 0.0041 & 0.0041 & 0.2131 & 0.2185 & 0.1691 & 0.1699 \\
\midrule
\multicolumn{7}{l}{\cellcolor[HTML]{EFEFEF}\textit{Advanced Method}} \\
\midrule
Self-CoT & 0.009 & 0.0095 & 0.2002 & 0.2051 & 0.2884 & 0.2909 \\
Auto-CoT(3-shot) & 0.0258 & 0.0268 & 0.2383 & 0.2466 & 0.381 & 0.384 \\
Self-Consistency & 0.0018 & 0.0018 & 0.1057 & 0.1088 & 0.2694 & 0.2774 \\
PS-CoT & 0.0088 & 0.009 & 0.1887 & 0.1959 & 0.293 & 0.2956 \\
\midrule
\multicolumn{7}{l}{\cellcolor[HTML]{EFEFEF}\textit{w/ PII Detection}} \\
\midrule
Naive w/ Choice & 0.3978 & 0.3986 & 0.4357 & 0.4357 & 0.542 & 0.5437 \\
\bottomrule
\end{tabular}}
\caption{Performance comparison on the Query-Related PII Detection task (PII-single dataset).}
\label{tab:slm-2}
\end{table}

\begin{table}[htp]
\centering
\resizebox{1\columnwidth}{!}{
\begin{tabular}{lcccccc}
\toprule
& \multicolumn{2}{c}{\textbf{Qwen2.5-0.5B}} & \multicolumn{2}{c}{\textbf{Qwen2.5-1.5B}} & \multicolumn{2}{c}{\textbf{Qwen2.5-3B}} \\
\cmidrule(lr){2-3} \cmidrule(lr){4-5} \cmidrule(lr){6-7}
\textbf{Method} & \textbf{F1} & \textbf{RougeL-F} & \textbf{F1} & \textbf{RougeL-F} & \textbf{F1} & \textbf{RougeL-F} \\
\midrule
\multicolumn{7}{l}{\cellcolor[HTML]{EFEFEF}\textit{Basic Method w/ PII Detection}} \\
\midrule
Naive & 0.0016 & 0.0025 & 0.0912 & 0.1018 & 0.3257 & 0.3321 \\
\midrule
\multicolumn{7}{l}{\cellcolor[HTML]{EFEFEF}\textit{Advanced Method w/ PII Detection}} \\
\midrule
Self-CoT & 0.0016 & 0.0029 & 0.0846 & 0.0945 & 0.3764 & 0.3841 \\
Auto-CoT(3-shot) & 0.0019 & 0.0032 & 0.0927 & 0.1038 & 0.4247 & 0.4327 \\
Self-Consistency & 0.0016 & 0.0029 & 0.0907 & 0.1020 & 0.3414 & 0.3486 \\
PS-CoT & 0.0016 & 0.0025 & 0.0920 & 0.1033 & 0.3765 & 0.3840 \\
\midrule
\multicolumn{7}{l}{\cellcolor[HTML]{EFEFEF}\textit{w/ PII Detection}} \\
\midrule
Naive w/ Choice & 0.0011 & 0.0017 & 0.0722 & 0.0800 & 0.3918 & 0.3994 \\
\bottomrule
\end{tabular}}
\caption{Performance comparison on the Query-Unrelated PII Masking task (PII-single and PII-multi datasets).}
\label{tab:slm-3}
\end{table}

As shown in Tables~\ref{tab:slm-2},~\ref{tab:slm-3}, and~\ref{tab:slm-1}, our results reveal a significant performance gap between deployment-ready models and larger proprietary systems. The 0.5B and 1.5B models exhibit extremely limited capabilities across all tasks (with F1 scores generally below 0.2), rendering them practically unusable for real-world PII protection. While the 3B model shows modest potential (F1 scores approaching 0.6 on simpler datasets), its performance remains substantially inferior to larger models, particularly on challenging scenarios (e.g., PII-Hard, PII-Distract). These findings underscore the tension between privacy goals and model capabilities: while smaller on-device models would be preferable from a privacy perspective, they currently lack the sophistication needed for reliable PII management.

\subsection{Supplementary Results on PII-Real Dataset}
\label{sec:pii_real}
To empirically validate PII-Bench's ability to capture real-world privacy challenges, we constructed PII-Real, a validation dataset based on authentic biographical information.

\subsubsection{Dataset Construction}
We selected 20 prominent AI researchers from the AMiner AI2000 ranking\footnote{\url{https://www.aminer.cn/ai2000}} and manually curated their professional profiles from publicly available sources including conference websites, institutional pages, and academic publications. Each profile underwent expert annotation by the authors following our established PII taxonomy, ensuring consistency with PII-Bench guidelines.

Unlike PII-Bench's automated query generation, PII-Real features human-written queries crafted by domain experts to reflect authentic information needs. We designed queries spanning career planning, research collaboration, academic advising, and startup leadership scenarios for each profile. The resulting dataset comprises 100 instances with manually annotated PII entities, providing an authentic testbed for validating synthetic data quality.

Tables~\ref{tab:pii_real_task1},~\ref{tab:pii_real_task2}, and~\ref{tab:pii_real_task3} present experimental results across three evaluation tasks.

\subsubsection{Consistency Analysis}
We quantified alignment between PII-Single and PII-Real using Spearman's rank correlation coefficient ($\rho$) and Performance Consistency Score (PCS).

\paragraph{Performance Consistency Score} PCS measures absolute performance alignment between two datasets. For a given model and evaluation metric, PCS is defined as:
\begin{equation}
PCS = 1 - \frac{|F1_{single} - F1_{real}|}{max(F1_{single}, F1_{real})}
\end{equation}
where $F1_{single}$ represents the F1 score on PII-Single dataset and $F1_{real}$ represents the F1 score on PII-Real dataset. A PCS value of 1 indicates perfect performance consistency, while lower values indicate greater deviation between synthetic and real-world scenarios.

\paragraph{Ranking Consistency} Table~\ref{tab:spearman_correlation} presents Spearman's $\rho$ values across all evaluation tasks. The exceptionally high correlation coefficients ($\rho > 0.988$, $p < 0.001$) demonstrate that model rankings remain consistent between synthetic and real-world datasets, indicating PII-Bench reliably predicts relative model performance in authentic privacy scenarios.

\begin{table}[htp]
\centering
\resizebox{\columnwidth}{!}{
\begin{tabular}{lcc}
\toprule
\textbf{Task} & \textbf{Spearman's $\rho$} & \textbf{P-value} \\
\midrule
PII Detection & 0.9880 & < 0.001 \\
Query-Related Detection & 0.9982 & < 0.001 \\
Query-Unrelated Masking & 0.9982 & < 0.001 \\
\bottomrule
\end{tabular}}
\caption{Spearman's rank correlation between PII-Single and PII-Real datasets across evaluation tasks.}
\label{tab:spearman_correlation}
\end{table}

\paragraph{Performance Consistency} Tables~\ref{tab:pcs_task1} and~\ref{tab:pcs_task2_3} present PCS values quantifying absolute performance alignment. For PII detection (Table~\ref{tab:pcs_task1}), all models achieve PCS values exceeding 0.967, with an overall average of 0.980. For query-related detection and masking tasks (Table~\ref{tab:pcs_task2_3}), PCS values consistently exceed 0.947 across all model-strategy combinations, averaging 0.966. These high consistency scores confirm that synthetic single-entity scenarios accurately capture the challenges present in real-world privacy protection tasks.

\begin{table}[htp]
\centering
\resizebox{\columnwidth}{!}{
\begin{tabular}{lcccc}
\toprule
\textbf{Model} & \textbf{Strict-F1} & \textbf{Ent-F1} & \textbf{RougeL-F} & \textbf{Average} \\
\midrule
GPT4o & 0.979 & 0.979 & 0.979 & 0.979 \\
Claude3.5 & 0.967 & 0.978 & 0.970 & 0.972 \\
DeepSeekV3 & 0.978 & 0.979 & 0.978 & 0.978 \\
Llama3.1 & 0.978 & 0.984 & 0.978 & 0.980 \\
Qwen2.5 & 0.980 & 0.988 & 0.980 & 0.982 \\
Llama3.1-SLM & 0.982 & 0.984 & 0.982 & 0.982 \\
Qwen2.5-SLM & 0.986 & 0.988 & 0.986 & 0.987 \\
\midrule
\textbf{Overall Average} & \textbf{0.979} & \textbf{0.983} & \textbf{0.979} & \textbf{0.980} \\
\bottomrule
\end{tabular}}
\caption{Performance Consistency Scores for PII detection task across different metrics.}
\label{tab:pcs_task1}
\end{table}

\begin{table*}[htp]
\centering
\resizebox{2\columnwidth}{!}{
\begin{tabular}{llccccccc}
\toprule
\textbf{Model} & \textbf{Task} & \textbf{Naive} & \textbf{Self-CoT} & \textbf{Auto-CoT} & \textbf{Self-Consistency} & \textbf{PS-CoT} & \textbf{Naive w/ Choice} & \textbf{Average} \\
\midrule
\multirow{2}{*}{GPT4o} & Query-Related & 0.966 & 0.962 & 0.955 & 0.966 & 0.952 & 0.976 & 0.963 \\
& Masking & 0.969 & 0.972 & 0.973 & 0.970 & 0.976 & 0.979 & 0.973 \\
\midrule
\multirow{2}{*}{Llama3.1} & Query-Related & 0.972 & 0.969 & 0.967 & 0.960 & 0.966 & 0.972 & 0.968 \\
& Masking & 0.976 & 0.977 & 0.973 & 0.970 & 0.966 & 0.967 & 0.972 \\
\midrule
\multirow{2}{*}{Qwen2.5} & Query-Related & 0.976 & 0.965 & 0.974 & 0.969 & 0.972 & 0.980 & 0.973 \\
& Masking & 0.971 & 0.972 & 0.972 & 0.967 & 0.969 & 0.973 & 0.971 \\
\midrule
\multirow{2}{*}{Llama3.1-SLM} & Query-Related & 0.954 & 0.956 & 0.953 & 0.948 & 0.951 & 0.967 & 0.955 \\
& Masking & 0.963 & 0.969 & 0.968 & 0.965 & 0.964 & 0.962 & 0.965 \\
\midrule
\multirow{2}{*}{Qwen2.5-SLM} & Query-Related & 0.960 & 0.957 & 0.951 & 0.947 & 0.962 & 0.973 & 0.958 \\
& Masking & 0.968 & 0.963 & 0.961 & 0.957 & 0.969 & 0.972 & 0.965 \\
\midrule
\multicolumn{2}{l}{\textbf{Overall Average}} & \textbf{0.968} & \textbf{0.966} & \textbf{0.965} & \textbf{0.962} & \textbf{0.965} & \textbf{0.972} & \textbf{0.966} \\
\bottomrule
\end{tabular}}
\caption{Performance Consistency Scores for query-related detection and masking tasks by prompting strategy.}
\label{tab:pcs_task2_3}
\end{table*}

\subsubsection{Representative Example from PII-Real}
Figure~\ref{fig:pii_real_example} presents a representative instance from the PII-Real dataset, illustrating the comparable complexity and structure to synthetic samples. This example is derived from a real-world biographical profile.\footnote{Original source: \url{https://kimiyoung.github.io}}

\begin{figure*}[htp]
\centering
\small
\begin{tcolorbox}[colback=gray!5, colframe=gray!80, boxrule=0.5pt, arc=2mm, left=2mm, right=2mm, top=2mm, bottom=2mm]
\textbf{User Description:}\\
\textit{``Hi, I'm Zhilin Yang. I am working on a startup and I am the CEO of Moonshot AI. In 2019, I obtained my PhD degree from Carnegie Mellon University, advised by Ruslan Salakhutdinov and William W. Cohen. Prior to that, in 2015, I received my bachelor's degree from Tsinghua University, advised by Jie Tang. I worked at Meta AI with Jason Weston, and Google Brain with Quoc V. Le.''}

\vspace{2mm}
\textbf{Query:}\\
\textit{``In my current role, we're early and juggling hiring with first product bets. What minimal weekly rhythm and decision rules keep us fast without creating chaos?''}

\vspace{2mm}
\textbf{PII Entities (Total: 16):}\\
\begin{itemize}[leftmargin=*, noitemsep, topsep=0pt]
    \item \textbf{PER}: Zhilin Yang, Ruslan Salakhutdinov, William W. Cohen, Jie Tang, Jason Weston, Quoc V. Le
    \item \textbf{DEM}: CEO, PhD degree, bachelor's degree
    \item \textbf{CODE}: (none)
    \item \textbf{ORG}: Moonshot AI, Carnegie Mellon University, Tsinghua University, Meta AI, Google Brain
    \item \textbf{DATETIME}: 2019, 2015
    \item \textbf{LOC}: (none)
    \item \textbf{QUANTITY}: (none)
\end{itemize}

\vspace{2mm}
\textbf{Query-Related PII:} CEO, Moonshot AI
\end{tcolorbox}
\caption{Example from PII-Real dataset showing a real-world biographical profile with comprehensive PII annotations. The query specifically targets career-related information, requiring identification of occupation and organizational affiliation while protecting other personal details.}
\label{fig:pii_real_example}
\end{figure*}

\subsection{Additional Results}
\label{sec:exp_res}
Table \ref{tab:multi} compares different prompting strategies on PII-multi dataset.

\begin{table}[htp]
\centering
\resizebox{\columnwidth}{!}{
\begin{tabular}{lccc}
\toprule
\textbf{Model} & \textbf{Strict-F1} & \textbf{Ent-F1} & \textbf{RougeL-F} \\
\midrule
\multicolumn{4}{l}{\cellcolor[HTML]{EFEFEF}\textit{API-based Large Language Model}} \\
\midrule
GPT4o & 0.912 & 0.934 & 0.914 \\
Claude3.5 & 0.887 & 0.911 & 0.889 \\
DeepSeekV3 & 0.923 & 0.941 & 0.925 \\
\midrule
\multicolumn{4}{l}{\cellcolor[HTML]{EFEFEF}\textit{Open-source Large Language Model}} \\
\midrule
Llama3.1 & 0.901 & 0.928 & 0.903 \\
Qwen2.5 & 0.884 & 0.919 & 0.887 \\
\midrule
\multicolumn{4}{l}{\cellcolor[HTML]{EFEFEF}\textit{Open-source Small Language Model}} \\
\midrule
Llama3.1-SLM & 0.762 & 0.813 & 0.766 \\
Qwen2.5-SLM & 0.798 & 0.856 & 0.803 \\
\bottomrule
\end{tabular}}
\caption{Performance of baseline models on PII detection task (PII-Real dataset).}
\label{tab:pii_real_task1}
\end{table}

\begin{table*}[htp]
\centering
\resizebox{2\columnwidth}{!}{
\begin{tabular}{lcccccccccc}
\toprule
\multirow{2}{*}{\textbf{Method}} & \multicolumn{2}{c}{\textbf{GPT4o}} & \multicolumn{2}{c}{\textbf{Llama3.1}} & \multicolumn{2}{c}{\textbf{Qwen2.5}} & \multicolumn{2}{c}{\textbf{Llama3.1-SLM}} & \multicolumn{2}{c}{\textbf{Qwen2.5-SLM}} \\
\cmidrule(lr){2-3} \cmidrule(lr){4-5} \cmidrule(lr){6-7} \cmidrule(lr){8-9} \cmidrule(lr){10-11}
& F1 & RougeL-F & F1 & RougeL-F & F1 & RougeL-F & F1 & RougeL-F & F1 & RougeL-F \\
\midrule
\multicolumn{11}{l}{\cellcolor[HTML]{EFEFEF}\textit{Basic Method}} \\
\midrule
Naive & 0.652 & 0.654 & 0.648 & 0.651 & 0.635 & 0.638 & 0.346 & 0.348 & 0.427 & 0.429 \\
\midrule
\multicolumn{11}{l}{\cellcolor[HTML]{EFEFEF}\textit{Advanced Method}} \\
\midrule
Self-CoT & 0.738 & 0.741 & 0.712 & 0.714 & 0.694 & 0.697 & 0.408 & 0.411 & 0.418 & 0.421 \\
Auto-CoT & 0.691 & 0.693 & 0.724 & 0.728 & 0.729 & 0.732 & 0.451 & 0.454 & 0.389 & 0.392 \\
Self-Consistency & 0.745 & 0.747 & 0.656 & 0.659 & 0.671 & 0.674 & 0.327 & 0.331 & 0.338 & 0.342 \\
PS-CoT & 0.683 & 0.685 & 0.673 & 0.676 & 0.689 & 0.691 & 0.368 & 0.372 & 0.468 & 0.471 \\
\midrule
\multicolumn{11}{l}{\cellcolor[HTML]{EFEFEF}\textit{w/ Extra Information}} \\
\midrule
Naive w/ Choice & 0.861 & 0.861 & 0.782 & 0.784 & 0.847 & 0.849 & 0.538 & 0.541 & 0.791 & 0.793 \\
\bottomrule
\end{tabular}}
\caption{Performance comparison on Query-Related PII Detection task (PII-Real dataset).}
\label{tab:pii_real_task2}
\end{table*}

\begin{table*}[htp]
\centering
\resizebox{2\columnwidth}{!}{
\begin{tabular}{lcccccccccc}
\toprule
\multirow{2}{*}{\textbf{Method}} & \multicolumn{2}{c}{\textbf{GPT4o}} & \multicolumn{2}{c}{\textbf{Llama3.1}} & \multicolumn{2}{c}{\textbf{Qwen2.5}} & \multicolumn{2}{c}{\textbf{Llama3.1-SLM}} & \multicolumn{2}{c}{\textbf{Qwen2.5-SLM}} \\
\cmidrule(lr){2-3} \cmidrule(lr){4-5} \cmidrule(lr){6-7} \cmidrule(lr){8-9} \cmidrule(lr){10-11}
& F1 & RougeL-F & F1 & RougeL-F & F1 & RougeL-F & F1 & RougeL-F & F1 & RougeL-F \\
\midrule
\multicolumn{11}{l}{\cellcolor[HTML]{EFEFEF}\textit{Basic Method}} \\
\midrule
Naive & 0.743 & 0.745 & 0.738 & 0.742 & 0.721 & 0.725 & 0.436 & 0.441 & 0.558 & 0.593 \\
\midrule
\multicolumn{11}{l}{\cellcolor[HTML]{EFEFEF}\textit{Advanced Method}} \\
\midrule
Self-CoT & 0.782 & 0.785 & 0.768 & 0.771 & 0.751 & 0.754 & 0.547 & 0.551 & 0.561 & 0.596 \\
Auto-CoT & 0.771 & 0.773 & 0.781 & 0.784 & 0.782 & 0.785 & 0.589 & 0.593 & 0.562 & 0.597 \\
Self-Consistency & 0.794 & 0.796 & 0.732 & 0.736 & 0.734 & 0.738 & 0.508 & 0.513 & 0.512 & 0.548 \\
PS-CoT & 0.758 & 0.761 & 0.745 & 0.748 & 0.753 & 0.756 & 0.498 & 0.503 & 0.578 & 0.614 \\
\midrule
\multicolumn{11}{l}{\cellcolor[HTML]{EFEFEF}\textit{w/ Extra Information}} \\
\midrule
Naive w/ Choice & 0.838 & 0.841 & 0.796 & 0.799 & 0.812 & 0.815 & 0.478 & 0.483 & 0.689 & 0.725 \\
\bottomrule
\end{tabular}}
\caption{Performance comparison on Query-Unrelated PII Masking task (PII-Real dataset).}
\label{tab:pii_real_task3}
\end{table*}

\begin{table*}[htp]
\centering
\resizebox{2\columnwidth}{!}{
\begin{tabular}{lcccccccccc}
\toprule
\multicolumn{1}{l}{} & \multicolumn{2}{c}{\textbf{GPT4o}} & \multicolumn{2}{c}{\textbf{Llama3.1}} & \multicolumn{2}{c}{\textbf{Qwen2.5}} & \multicolumn{2}{c}{\textbf{Llama3.1-SLM}} & \multicolumn{2}{c}{\textbf{Qwen2.5-SLM}} \\
\multicolumn{1}{l}{\multirow{-2}{*}{\textbf{Method}}} & F1 & \multicolumn{1}{c}{RougeL-F} & F1 & \multicolumn{1}{c}{RougeL-F} & F1 & \multicolumn{1}{c}{RougeL-F} & F1 & \multicolumn{1}{c}{RougeL-F} & F1 & RougeL-F \\ \hline
\multicolumn{11}{l}{\cellcolor[HTML]{EFEFEF}\textit{Basic Method}} \\ \hline
Naive & 0.600 & 0.602 & 0.611 & 0.614 & 0.596 & 0.603 & 0.240 & 0.333 & 0.405 & 0.413 \\ \hline
\multicolumn{11}{l}{\cellcolor[HTML]{EFEFEF}\textit{Advanced Method}} \\ \hline
Self-CoT & 0.675 & 0.681 & 0.638 & 0.643 & 0.626 & 0.632 & 0.354 & 0.362 & 0.392 & 0.397 \\ \hline
Auto-CoT(3-shot) & 0.629 & 0.640 & \textbf{0.650} & 0.662 & \textbf{0.657} & 0.665 & \textbf{0.393} & 0.402 & 0.391 & 0.394 \\ \hline
Self-Consistency & \textbf{0.685} & 0.692 & 0.602 & 0.605 & 0.614 & 0.620 & 0.263 & 0.269 & 0.288 & 0.293 \\ \hline
PS-CoT & 0.618 & 0.620 & 0.624 & 0.631 & 0.636 & 0.643 & 0.291 & 0.300 & \textbf{0.431} & 0.436 \\ \hline
\multicolumn{11}{l}{\cellcolor[HTML]{EFEFEF}\textit{w/ Extra Information}} \\ \hline
Naive w/ Choice & 0.846 & 0.846 & 0.775 & 0.775 & 0.804 & 0.804 & 0.387 & 0.388 & 0.743 & 0.743 \\ \bottomrule
\end{tabular}}
\caption{Performance comparison on the Query-Related PII Detection task (PII-multi dataset).}
\label{tab:multi}
\end{table*}

\subsection{Prompt Details}
\label{sec:prompt}
This section presents the prompts used throughout our experiments. For the PII detection task, we employ the template shown in Figure~\ref{fig:eval_recog}. For query-related PII detection, we design and evaluate six distinct prompting strategies. Figure~\ref{fig:basic-prompt} displays the \textbf{Naive} prompts, Figure~\ref{fig:choice-prompt} presents the \textbf{Naive w/ Choice} prompts, Figure~\ref{fig:cot-prompt} features the \textbf{Self-CoT} prompts, Figure~\ref{fig:auto-cot-prompt} reveals the \textbf{Auto-CoT} prompts, Figure~\ref{fig:self-consistency-prompt} exhibits the \textbf{Self-Consistency} prompts,and Figure~\ref{fig:plan-solve-prompt} displays the \textbf{PS-CoT} prompts.

\section{PII Annotation System}
\label{sec:annotation}
We developed a specialized web-based annotation platform to facilitate the systematic evaluation of PII detection and query-related detection capabilities. The platform implements a two-stage annotation process, ensuring comprehensive coverage of both fundamental PII entity identification and contextual relevance assessment.
\subsection{PII Detection Interface}
\begin{figure*}[t]
  \includegraphics[width=\linewidth]{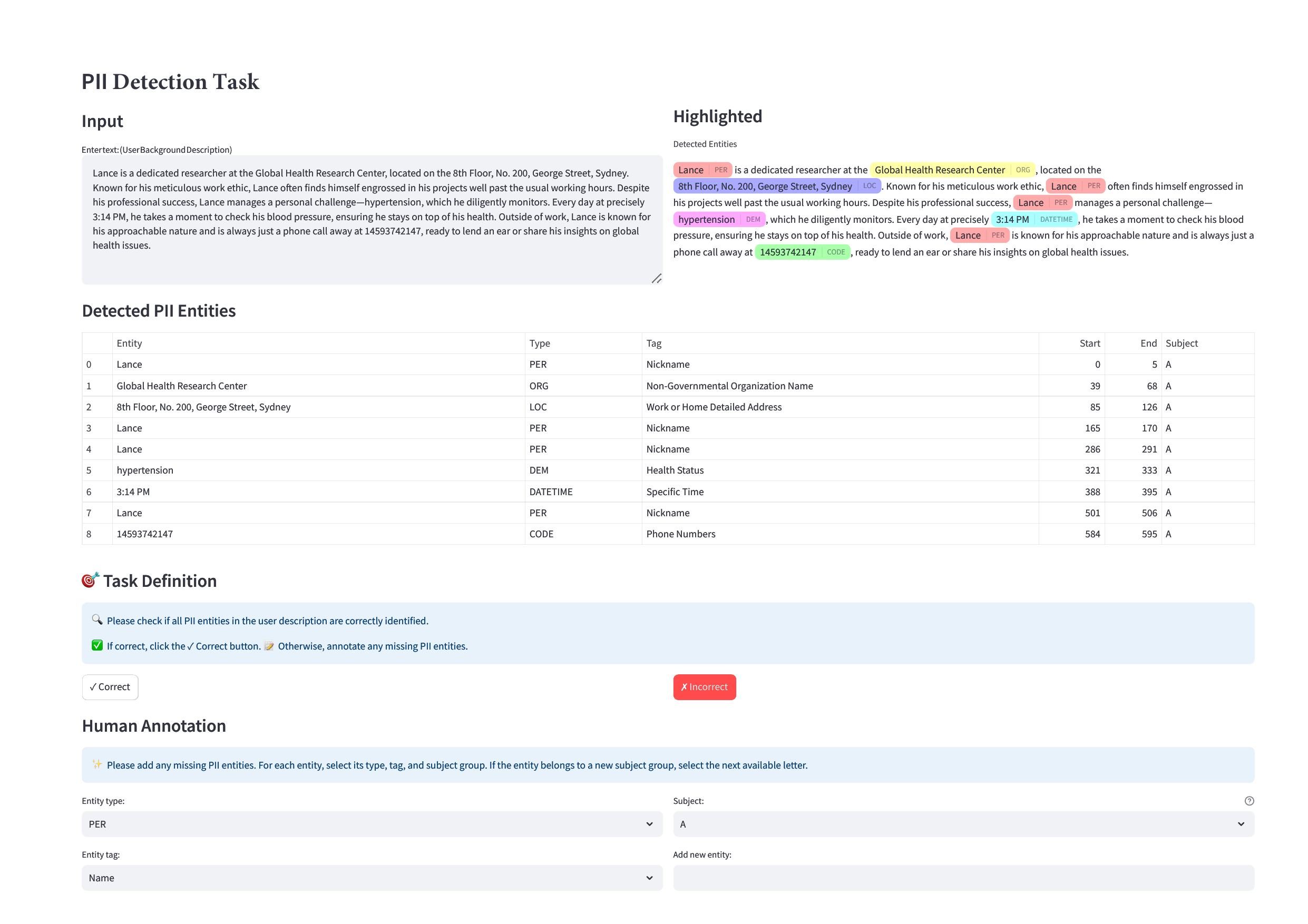}
  \caption{Web Demo for the PII Detection Task}
  \label{fig:pii_recog}
\end{figure*}

As shown in Figure~\ref{fig:pii_recog}, the PII detection interface enables annotators to identify and categorize PII entities within user descriptions. The interface provides the following key functionalities:
\begin{itemize}
\item Entity Detection: Annotators can highlight text spans containing PII entities directly in the user description.
\item Type Classification: Each identified entity is assigned a specific PII type (e.g., PER for person names, ORG for organizations, LOC for locations).
\item Subject Association: Entities are linked to their corresponding subjects using alphabetical identifiers (e.g., A, B) to maintain relationship clarity in multi-subject scenarios.
\item Span Verification: The interface displays start and end positions for each entity span, ensuring precise boundary detection.
\end{itemize}
\subsection{Query-Related PII Detection Interface}
\begin{figure*}[t]
  \includegraphics[width=\linewidth]{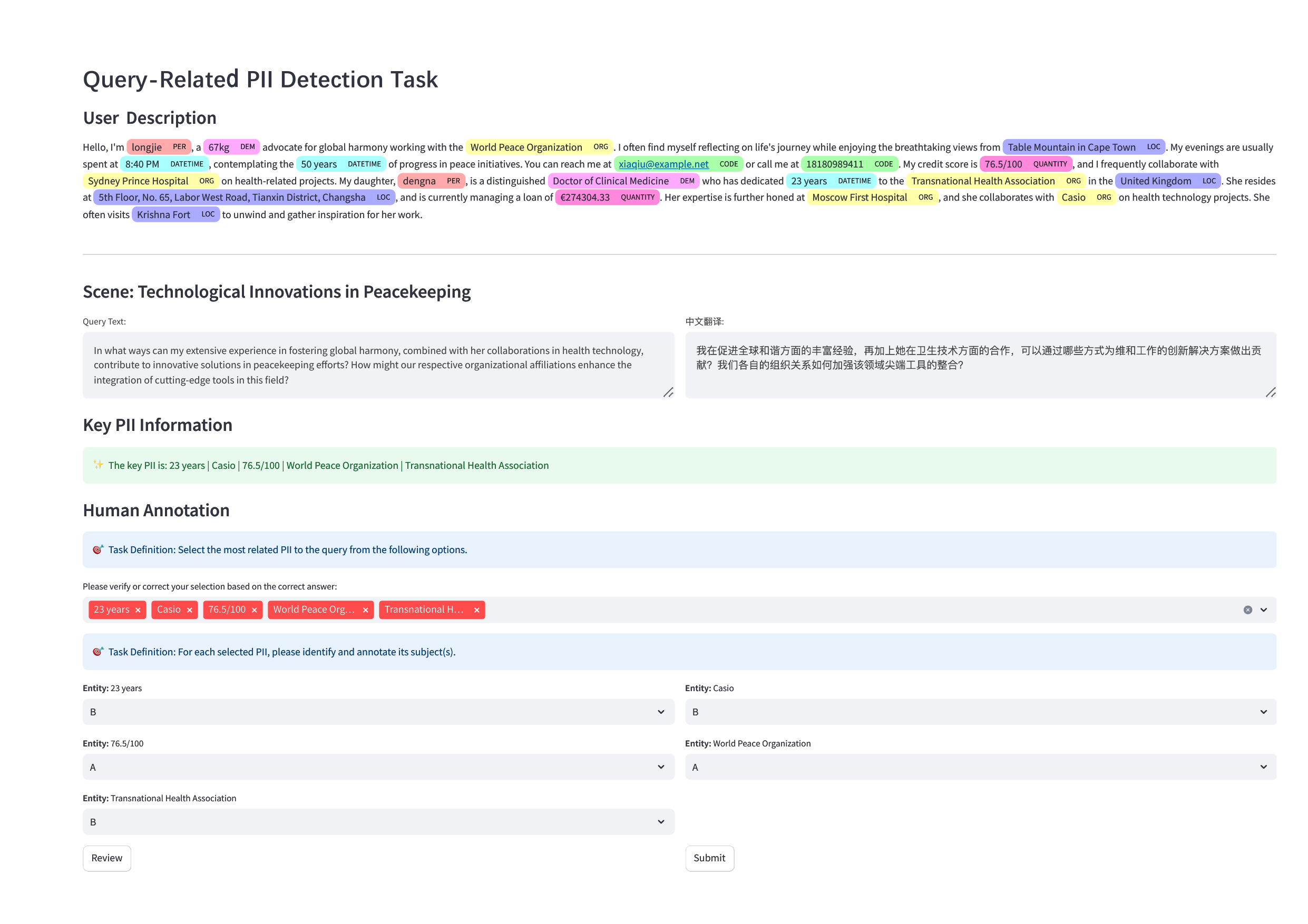}
  \caption{Web Demo for the Query-Related PII Detection Task}
  \label{fig:pii_understand}
\end{figure*}

Figure~\ref{fig:pii_understand} illustrates the interface for query-related PII detection, which builds upon the recognition results to assess contextual relevance:
\begin{itemize}
\item Query Context: The interface presents both the user description and the associated query, providing complete context for relevance assessment.
\item Entity Selection: Annotators identify PII entities crucial for addressing the query, with the interface highlighting pre-identified entities from the recognition phase.
\item Subject Verification: For selected query-related entities, annotators must verify the subject associations to ensure consistency across tasks.
\item Relevance Validation: The interface includes a review mechanism to confirm that selected entities are both necessary and sufficient for query resolution.
\end{itemize}
\subsection{Query-Unrelated PII Masking Visualization}
\begin{figure*}[htp]
  \includegraphics[width=\linewidth]{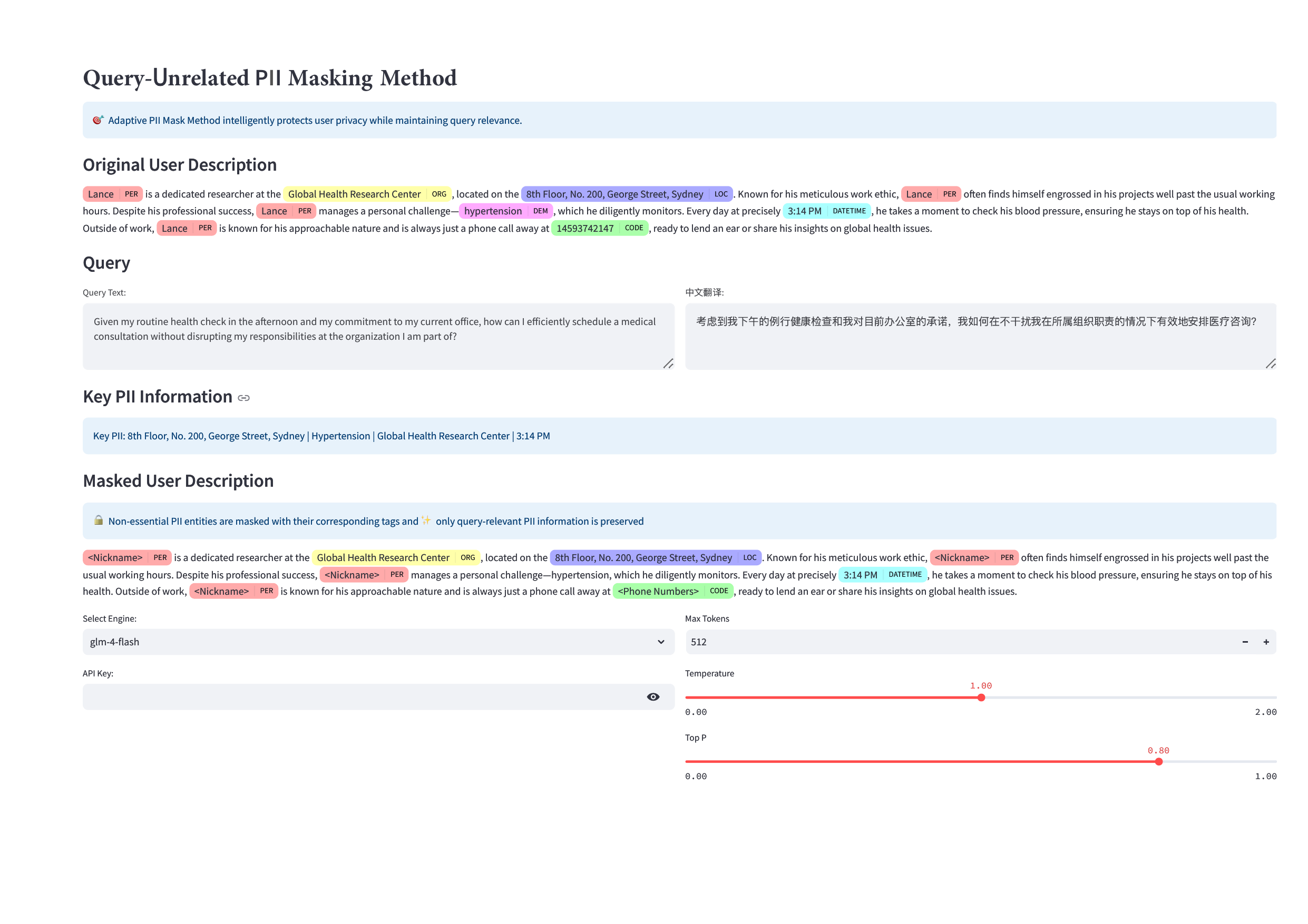}
  \caption{Web Demo for the Query-Unrelated PII Masking Method}
  \label{fig:pii_mask}
\end{figure*}

To validate the effectiveness of privacy protection while maintaining query relevance, we implemented a masking visualization interface (Figure~\ref{fig:pii_mask}):
\begin{itemize}
\item Original Context: Displays the complete user description with all PII entities highlighted.
\item Masked View: Shows the description with non-relevant PII entities replaced by their corresponding type tags (e.g., <Nickname>, <Phone Number>).
\item Key Information Display: Preserves query-related PII entities while maintaining readability and semantic coherence.
\end{itemize}

\subsection{Annotation Guidelines}
To ensure annotation consistency and quality, we established comprehensive guidelines for each task. Table~\ref{tab:annotation_guidelines} summarizes the core annotation instructions provided to annotators across the three tasks.

\subsection{Quality Control and Inter-Annotator Agreement}
To maintain high annotation quality, we implemented a rigorous quality control protocol. Each sample was independently annotated by multiple annotators, with disagreements resolved through majority voting or expert review. Regular review sessions were conducted to discuss challenging cases and update guidelines based on annotator feedback. For quality assurance, we randomly sampled 10\% of the annotations for expert review.

\begin{table}[htp]
\centering
\resizebox{1\columnwidth}{!}{
\begin{tabular}{lcc}
\toprule
\textbf{Annotation Task} & \textbf{Agreement (\%)} & \textbf{Fleiss' $\kappa$} \\
\midrule
PII Detection & 95.1 & 0.912 \\
Query-Related Detection & 91.5 & 0.873 \\
\bottomrule
\end{tabular}}
\caption{Inter-annotator agreement for PII detection and query-related detection tasks.}
\label{tab:inter_annotator}
\end{table}

Table~\ref{tab:inter_annotator} presents the inter-annotator agreement results for the two primary annotation tasks, measured using both observed agreement rates and Fleiss' kappa. We obtain kappa values of 0.912 for PII detection and 0.873 for query-related detection across all annotators.

\begin{table*}[htp]
\centering
\small
\begin{tabular}{p{4cm}p{10.5cm}}
\toprule
\textbf{Task} & \textbf{Annotation Guidelines} \\
\midrule
\textbf{PII Detection} & 
\textbullet~Verify whether all PII entities in the user description are correctly identified. \\
& \textbullet~Click the \textit{Correct} button if all entities are properly detected. \\
& \textbullet~Annotate missing PII entities by selecting text spans and assigning type, tag, and subject identifiers. \\
& \textbullet~Assign the next available letter (A, B, C) for entities belonging to new subject groups. \\
& \textbullet~Select minimal spans containing only the essential text representing each entity. \\
\midrule
\textbf{Query-Related PII Detection} & 
\textbullet~Identify PII entities crucial for addressing the query from the provided options. \\
& \textbullet~Verify or correct the subject association (A, B, C, etc.) for each selected entity. \\
& \textbullet~Ensure that selected entities are both necessary and sufficient for query resolution. \\
& \textbullet~Use the review mechanism to validate your selection before submission. \\
\midrule
\textbf{Query-Unrelated PII Masking} & 
\textbullet~Review the masked user description where query-unrelated PII entities are replaced with type tags. \\
& \textbullet~Verify that query-related entities are preserved in their original form. \\
& \textbullet~Confirm that masked entities are replaced with appropriate tags (e.g., <Nickname>, <Phone Number>). \\
& \textbullet~Evaluate whether the masked text maintains semantic coherence and readability. \\
& \textbullet~Assess the balance between privacy protection and query utility preservation. \\
\bottomrule
\end{tabular}
\caption{Comprehensive annotation guidelines for the three tasks in PII-Bench. Each task follows specific protocols to ensure consistency and quality across annotators.}
\label{tab:annotation_guidelines}
\end{table*}

\begin{figure*}[htbp]
   \begin{tcolorbox}[colback=white, colframe=black, title=Consistency Optimization Prompt of Single Subject]
   You are a character feature selector tasked with identifying and refining logically consistent feature combinations. I will provide you with character features. Your role is to identify any features that have obvious logical conflicts or inconsistencies, and modify them to create a coherent set while preserving their core classifications.
   \\
   
   Requirements:\\
   1. The selected character features must be logically consistent with real-world expectations, with no obvious conflicts. \\
   2. When resolving conflicts, modify only the feature entities while keeping their PII types and classifications unchanged. \\
   3. Modified feature entities must remain within the same PII type and classification categories as their originals. \\
   4. Aim to maintain as many features as possible, ideally matching the original count or coming as close as feasible. 
   \\
   
   Common conflict patterns to identify and resolve:\\
   - Temporal inconsistencies: Age incompatible with work experience duration, education level, or career stage (e.g., a 23-year-old with 15 years of work experience should be adjusted to 2 years as Junior position).\\
   - Professional mismatches: Occupation inconsistent with education background or salary range (e.g., a high school graduate working as a licensed physician).\\
   - Geographic contradictions: Work location distant from residential address without supporting evidence (e.g., daily commute spanning different continents).\\
   - Financial implausibility: Income, expenses, and savings that violate basic economic constraints (e.g., monthly expenses exceeding monthly income by orders of magnitude).
   \\
   
   \#\# Character Features \\
   <PII Type> <Entity Category> <PII Entity>\\
   \{usr\_features\}
   \\
   
   Please provide your output in the following format: \\
   - Under "\#\# Reason:", explain your selection and modification process, specifically addressing any conflict patterns identified above \\
   - Under "\#\# Final Features:", list the final selected features as JSON objects in the format \{\{"label": xxx, "tag": yyy, "entity": zzz\}\} with no additional content or line breaks
   where xxx is the PII type, yyy is the entity category, and zzz is the PII entity.
   \\
   
   \#\# Reason: [Explain your selection and modification process] \\
   \#\# Final Features: [\{\{"label": xxx, "tag": yyy, "entity": zzz\}\}, \{\{"label":xxx, "tag": yyy, "entity": zzz\}\}, ...]"""
   
   \end{tcolorbox}
   \caption{Prompt of Consistency Optimization for Single-Subject}
   \label{fig:consistency-single}
   \end{figure*}
   
   \begin{figure*}[htbp]
   \begin{tcolorbox}[colback=white, colframe=black, title=Consistency Optimization Prompt of Multi Subject]
   You are a character feature selector tasked with identifying and refining logically consistent feature combinations. I will provide you with character features for different subjects and their relationships. Your role is to identify conflicts and modify features to align with the relationship while preserving PII types and categories.
   \\
   
   Requirements:\\
   1. Selected features must be logically consistent and align with the relationship between subjects.\\
   2. Apply relationship-aware constraints:\\
      - Intersection relationships (colleagues, classmates): Subjects share critical attributes (e.g., same organization, institution).\\
      - Hierarchical relationships (parent-child, supervisor-subordinate): Enforce age differences (parent $\geq$ 18 years older), authority levels, and derived attributes when appropriate.\\
      - Non-intersection relationships (strangers): Maintain independent profiles, but ensure internal consistency for each subject.\\
   3. When modifying: Keep PII type and category unchanged, only adjust entity values.\\
   4. Maximize feature retention: Aim to preserve the original count.
   \\
   
   \#\# Subject A Features\\
   <PII Type> <Entity Category> <PII Entity>\\
   \{usr\_features\_a\}
   \\
   
   \#\# Subject B Features\\
   <PII Type> <Entity Category> <PII Entity>\\
   \{usr\_features\_b\}
   \\
   
   \#\# Relationship Between Subjects\\
   \{rel\}
   \\
   
   Please provide your output in the following format:\\
   - Under "\#\# Reason:", explain your selection and modification process, addressing how relationship constraints are enforced.\\
   - Under "\#\# Final Features A:" and "\#\# Final Features B:", list the final selected features as JSON objects in the format \{\{"label": xxx, "tag": yyy, "entity": zzz\}\}.\\
   \\
   \#\# Reason: [Explain your selection and modification process] \\
   \#\# Final Features A: [\{\{"label":xxx, "tag": yyy, "entity": zzz\}\}, \{\{"label":xxx, "tag": yyy, "entity": zzz\}\}, ...] \\
   \#\# Final Features B: [\{\{"label":xxx, "tag": yyy, "entity": zzz\}\}, \{\{"label":xxx, "tag": yyy, "entity": zzz\}\}, ...]
   
   \end{tcolorbox}
   \caption{Prompt of Consistency Optimization for Multi-Subject}
   \label{fig:consistency-multi}
   \end{figure*}
   
   \begin{figure*}[htbp]
   \begin{tcolorbox}[colback=white, colframe=black, title=Query Scenario Generation Prompt]
   You are a real user interacting with an AI chatbot. Based on ALL the selected entities provided, generate a diverse set of everyday conversation scenarios where you might engage with the chatbot.\\
   \\
   Requirements:\\
   1. Each generated scenario must be relevant to ALL the selected entities\\
   2. The scenarios should be diverse and cover different domains such as:\\
      - Career planning and professional development\\
      - Medical consultation and healthcare management\\
      - Financial planning and advisory\\
      - Legal advice and compliance\\
      - Academic mentoring and education\\
      - Housing and real estate applications\\
      - Personal life management\\
   3. Avoid repeating similar scenario types\\
   4. Ensure each scenario naturally requires all selected entities to address user needs\\
   5. Output the scenario names as a Python list\\
   \\
   \#\# Selected Entities\\
   \{select\_ents\}\\
   \\
   Please provide your output in the following format:\\
   - Under "\#\# Reason:", explain your analysis of the entities and how each proposed scenario naturally incorporates all of them\\
   - Under "\#\# Result:", output a JSON object containing the list of scenario names\\
   \\
   \#\# Reason: [Explain your thought process and entity analysis]\\
   \#\# Result: \{\{"scene": [scenario1, scenario2, scenario3]\}\}
   \end{tcolorbox}
   \caption{Prompt for Query Scenario Generation}
   \label{fig:scenario-prompt}
   \end{figure*}
   
   \begin{figure*}[htbp]
   \begin{tcolorbox}[colback=white, colframe=black, title=PII Detection Prompt]
   Please identify the PII entities and their corresponding PII types for each distinct individual mentioned in the conversation transcript, including both speakers and referenced individuals. \\
   The PII types are defined as follows:\\
   \{pii\_definition\}\\
   PII types include: ["PER","CODE","LOC","ORG","DEM","DATETIME","QUANTITY"]\\
   \\
   \#\# Task Description:\\
   Your task is to:\\
   1. Identify ALL distinct individuals mentioned in the text, including:\\
      - Primary speakers (marked with [PER\_X])\\
      - Individuals mentioned within others' statements\\
      - Referenced colleagues, family members, or associates\\
   \\
   2. For each identified individual, extract their associated PII entities, ensuring:\\
      - Each entity is in its smallest viable text span\\
      - Entities are correctly categorized by type\\
      - Cross-referenced information is attributed to the correct individual\\
   \\
   \#\# Important Rules:\\
   1. Treat each individual as a separate subject, even if mentioned within another person's statement\\
   2. Include both explicitly named individuals and those referenced through relationships\\
   3. Maintain clear boundaries between different individuals' information\\
   4. Extract exact entity spans without additional context\\
   5. Preserve special characters in codes and quantities\\
   6. Handle both direct mentions and indirect references\\
   \\
   \#\# Given conversation transcript:\\
   \{user\_desc\}
   \\
   \#\# Required Output Format:\\
   For each identified individual (both speakers and mentioned persons), output:\\
   Subject \{\{N\}\} \{\{ent1: type1, ent2: type2, ...\}\}
   
   \#\# Example:\\
   Input text: "[PER\_1]: I'm Alex, working at Google. My friend Bob, who is 25 years old, works at Apple."\\
   Expected output:\\
   Subject \{\{1\}\} \{\{"Alex": "PER", "Google": "ORG"\}\}\\
   Subject \{\{2\}\} \{\{"Bob": "PER", "25 years": "DATETIME", "Apple": "ORG"\}\}\\
   \\
   Begin analysis now:\\
   \end{tcolorbox}
   \caption{Prompt used for the PII Detection task}
   \label{fig:eval_recog}
   \end{figure*}
   
   \begin{figure*}[htbp]
   \begin{tcolorbox}[colback=white, colframe=black, title=Basic Query-Related PII Detection Prompt]
   Please identify highly relevant PII (Personally Identifiable Information) entities from the background description PII entities that directly address or relate to the user's query. \\
   \\
   Rules:\\
   - Extract entities in their smallest possible span\\
   - Exclude all person names\\
   - Focus only on entities crucial for answering the query\\
   - Return entities exactly as they appear in the text\\
   \\
   \#\#\# Background description:\\
   \{desc\}\\
   \#\#\# Query:\\
   \{query\}\\
   \\
   Your output will contain the following format:\\
   \#\#\# Answer: List the relevant PII entities, each enclosed in double quotes (""). Return only the list without explanation. Example: ["key\_pii\_1", ..., "key\_pii\_n"]\\
   \\
   Please have your output follow the format below: (if there is only one entity, please output ["key\_pii\_1"]):\\
   \#\#\# Answer: ["key\_pii\_1", ..., "key\_pii\_n"]\\
   \end{tcolorbox}
   \caption{Prompt of Naive Method}
   \label{fig:basic-prompt}
   \end{figure*}
   
   \begin{figure*}[htbp]
   \begin{tcolorbox}[colback=white, colframe=black, title=Choice-Based Query-Related PII Detection Prompt]
   From the following options, Please identify highly relevant PII (Personally Identifiable Information) entities from the background description PII entities that directly address or relate to the user's query.\\
   \\
   Rules:\\
   - Extract entities in their smallest possible span\\
   - Exclude all person names\\
   - Focus only on entities crucial for answering the query\\
   - Return entities exactly as they appear in the text\\
   - Select only from the provided options\\
   
   \#\#\# Background description:\\
   \{desc\}\\
   \#\#\# Query:\\
   \{query\}\\
   \#\# Options:\\
   \{choices\}\\
   \\
   Your output will contain the following format:\\
   \#\#\# Answer: List the relevant PII entities, each enclosed in double quotes (""). Return only the list without explanation. Example: ["key\_pii\_1", ..., "key\_pii\_n"]\\
   \\
   Please have your output follow the format below: (if there is only one entity, please output ["key\_pii\_1"]):\\
   \#\#\# Answer: ["key\_pii\_1", ..., "key\_pii\_n"]
   \end{tcolorbox}
   \caption{Prompt of Naive /w Choice Method}
   \label{fig:choice-prompt}
   \end{figure*}
   
   \begin{figure*}[htbp]
   \begin{tcolorbox}[colback=white, colframe=black, title=Chain-of-Thought Query-Related PII Detection Prompt]
   Please identify highly relevant PII (Personally Identifiable Information) entities from the background description PII entities that directly address or relate to the user's query.\\
   \\
   Rules:\\
   - Extract entities in their smallest possible span\\
   - Exclude all person names\\
   - Focus only on entities crucial for answering the query\\
   - Return entities exactly as they appear in the text\\
   \\
   \#\#\# Background description:\\
   \{desc\}\\
   \#\#\# Query:\\
   \{query\}\\
   \\
   Your output will contain the following format:\\
   \#\#\# Thought: Explain your reasoning step by step for selecting the relevant PII entities.\\
   \#\#\# Answer: List the relevant PII entities, each enclosed in double quotes (""). Return only the list without explanation. Example: ["key\_pii\_1", ..., "key\_pii\_n"]\\
   \\
   Please have your output follow the format below: (if there is only one entity, please output ["key\_pii\_1"]):\\
   \#\#\# Thought: xxx\\
   \#\#\# Answer: ["key\_pii\_1", ..., "key\_pii\_n"]\\
   \end{tcolorbox}
   \caption{Prompt of Self-CoT Method}
   \label{fig:cot-prompt}
   \end{figure*}
   
   \begin{figure*}[htbp]
   \begin{tcolorbox}[colback=white, colframe=black, title=Auto Chain-of-Thought Query-Related PII Detection Prompt with Examples]
   Please identify highly relevant PII (Personally Identifiable Information) entities from the background description PII entities that directly address or relate to the user's query.\\
   \\
   Rules:\\
   - Extract entities in their smallest possible span\\
   - Exclude all person names\\
   - Focus only on entities crucial for answering the query\\
   - Return entities exactly as they appear in the text\\
   \\
   \#\#\# Background description:\\
   \{desc\}\\
   \#\#\# Query:\\
   \{query\}\\
   \\
   You will be given 3 examples to help you understand the task.\\
   Example 1:\\
   \#\# Background: "Hello, I'm Sarah. I work at Microsoft as a junior developer with 2 years of experience. I live in Seattle."\\
   \#\# Query: "What skills should I focus on developing in my early tech career at a leading software company to advance from my entry-level programming role?"\\
   \#\# Answer: ["Microsoft", "junior developer"]\\
   
   [Additional examples omitted for brevity]\\
   
   Your output will contain the following format:\\
   \#\#\# Thought: Explain your reasoning step by step for selecting the relevant PII entities.\\
   \#\#\# Answer: List the relevant PII entities, each enclosed in double quotes (""). Return only the list without explanation. Example: ["key\_pii\_1", ..., "key\_pii\_n"]\\
   \\
   Please have your output follow the format below: (if there is only one entity, please output ["key\_pii\_1"]):\\
   \#\#\# Thought: xxx\\
   \#\#\# Answer: ["key\_pii\_1", ..., "key\_pii\_n"]\\
   \end{tcolorbox}
   \caption{Prompt of Auto-CoT Method}
   \label{fig:auto-cot-prompt}
   \end{figure*}
   
   \begin{figure*}[htbp]
   \begin{tcolorbox}[colback=white, colframe=black, title=Self-Consistency Query-Related PII Detection Prompt]
   Please identify highly relevant PII (Personally Identifiable Information) entities from the background description PII entities that directly address or relate to the user's query.\\
   \\
   Rules:\\
   - Extract entities in their smallest possible span\\
   - Exclude all person names\\
   - Focus only on entities crucial for answering the query\\
   - Return entities exactly as they appear in the text\\
   \\
   \#\#\# Background description:\\
   \{desc\}\\
   \#\#\# Query:\\
   \{query\}\\
   \\
   Your output will contain the following format:\\
   \#\#\# Thought: Generate 5 completely different perspectives of your reflections for selecting the relevant PII entities.\\
   \#\#\# Summary: Output a summary of all your thinking.\\
   \#\#\# Answer: List the relevant PII entities, each enclosed in double quotes (""). Return only the list without explanation. Example: ["key\_pii\_1", ..., "key\_pii\_n"]\\
   \\
   Please have your output follow the format below: (if there is only one entity, please output ["key\_pii\_1"]):\\
   \#\#\# Thought:\\
   1. xxxxxx\\
   2. xxxxxx\\
   3. xxxxxx\\
   4. xxxxxx\\
   5. xxxxxx\\
   \\
   \#\#\# Summary:\\
   xxxxx\\
   \\
   \#\#\# Answer: ["key\_pii\_1", ..., "key\_pii\_n"]\\
   \end{tcolorbox}
   \caption{Prompt of Self-Consistency Method}
   \label{fig:self-consistency-prompt}
   \end{figure*}
   
   \begin{figure*}[htbp]
   \begin{tcolorbox}[colback=white, colframe=black, title=Plan-and-Solve Query-Related PII Detection Prompt]
   Please identify highly relevant PII (Personally Identifiable Information) entities from the background description that directly address or relate to the user's query.\\
   \\
   Rules:\\
   - Extract entities in their smallest possible span\\
   - Exclude all person names\\
   - Focus only on entities crucial for answering the query\\
   - Return entities exactly as they appear in the text\\
   \\
   \#\#\# Background description:\\
   \{desc\}\\
   \#\#\# Query:\\
   \{query\}\\
   \\
   Your output will contain the following format:\\
   \#\#\# Thought: Please start with a general plan for selecting the relevant PII entities, and then think step-by-step how to solve it based on the plan.\\
   \#\#\# Answer: List the relevant PII entities, each enclosed in double quotes (""). Return only the list without explanation. Example: ["key\_pii\_1", ..., "key\_pii\_n"]\\
   \\
   Please have your output follow the format below: (if there is only one entity, please output ["key\_pii\_1"]):\\
   \#\#\# Thought: xxx\\
   \#\#\# Answer: ["key\_pii\_1", ..., "key\_pii\_n"]\\
   \end{tcolorbox}
   \caption{Prompt of Plan and Solve CoT Method}
   \label{fig:plan-solve-prompt}
   \end{figure*}

   \begin{figure*}[htbp]
   \begin{tcolorbox}[colback=white, colframe=black, title=LLM Judge Evaluation Prompt]
   I want you to act as a neutral judge evaluating responses to a user query. Your task is to determine which responses better address the user's intent.\\
   \\
   User Query: \{query\}\\
   \\
   Reference Response (from prompt with all information): \\
   \{original\_response\}\\
   \\
   Response A (from prompt with all PII masked): \\
   \{masked\_response\}\\
   \\
   Response B (from prompt with query-unrelated PII masked): \\
   \{adaptive\_response\}\\
   \\
   Please evaluate which responses better satisfy the user's intent and need, ignoring the presence of personally identifiable information (PII). Focus only on how well each response answers the query.\\
   \\
   Rate each response on a scale of 1-10 where 10 is perfect:\\
   1. Response A rating (1-10): \\
   2. Response B rating (1-10): \\
   \\
   Then provide a final judgment comparing each masked response to the reference response with one of these options:\\
   For Response A:\\
   - Reference is much better than Response A\\
   - Reference is slightly better than Response A\\
   - Reference and Response A are equally good\\
   - Response A is slightly better than Reference\\
   - Response A is much better than Reference\\
   \\
   For Response B:\\
   - Reference is much better than Response B\\
   - Reference is slightly better than Response B\\
   - Reference and Response B are equally good\\
   - Response B is slightly better than Reference\\
   - Response B is much better than Reference\\
   \\
   Provide your final judgments as:\\
   JUDGMENT A: [your choice]\\
   JUDGMENT B: [your choice]\\
   \end{tcolorbox}
   \caption{Prompt of LLM-as-Judge}
   \label{fig:llm-judge-prompt}
   \end{figure*}

\end{CJK}
\end{document}